Banner appropriate to article type will appear here in typeset article

# Wall-scaled eddies and embedded shear layers in high-Reynolds-number moderate adverse-pressure-gradient boundary layers

**Ahmad Zarei[1], Mitchell Lozier[1], Ivan Marusic[1,2] and Rahul Deshpande†[3]**

[1]Department of Mechanical Engineering, The University of Melbourne, Parkville, Victoria 3010, Australia
[2]Faculty of Mechanical Engineering and Naval Architecture, University of Zagreb, Croatia
[3]School of Engineering, RMIT University, Melbourne, Victoria 3000, Australia

This study compares the structure of high-Reynolds-number turbulent boundary layers (TBLs) under zero and low-to-moderate adverse pressure gradients (APGs), revealing consistent scaling behaviour and energy contributions associated with the wall-scaled (*i.e.*, attached) eddy hierarchy in both flows. The most pronounced differences arise in the outer/wake region, where the APG-induced energisation is shown to be closely linked to an outer-scaled, embedded-shear-layer-type flow organisation that progressively penetrates into the logarithmic region with increase in APG strength. These effects are investigated experimentally using two complementary datasets obtained for both ZPG and APG TBLs at matched friction Reynolds numbers ($\sim 10^4$) with minimal upstream PG history effects: a newly acquired two-point hot-wire dataset and a previously published two-dimensional particle image velocimetry (PIV) dataset. In the former experiment, one probe is fixed near the wall while the second traverses vertically across the entire TBL, enabling estimation of the spectral linear coherence spectrum (LCS). The resulting measurements confirm the geometric self-similarity of the wall-scaled eddy hierarchy that remains coherent with the wall. The LCS is subsequently employed as a spectral filter, revealing that the majority of the additional energy introduced by the APG is incoherent with the wall, and is linearly superimposed onto the wall-coherent component of the streamwise variance profile. This additional wall-incoherent energy is shown to underpin the departure of the streamwise variance profile from the inverse logarithmic law associated with the wall-coherent attached eddy hierarchy. Conditional averaging of the PIV datasets is performed to identify the flow structures associated with the APG-induced energy amplification. The analysis establishes direct links between the enhanced Reynolds stresses in the outer region of the APG TBL, the outer inflection point in the mean velocity profile, and an ejection–sweep organisation of the Reynolds shear stress, indicating strongly energised shear-layer-type motions that also exist, albeit weakly, in the ZPG TBL. Collectively, these findings reveal that pressure gradients primarily influence the competition between two coexisting dynamical pathways in high-Reynolds number TBLs: a wall-coherent pathway associated with the attached-eddy hierarchy and superstructures, and a wall-incoherent pathway associated with outer-scaled shear-layer-type motions, thereby providing a physical framework for future data-driven predictive models.

† Email address for correspondence: raadeshpande@gmail.com

**Abstract must not spill onto p.2**

## 1. Introduction and motivation

The structure and scaling of zero-pressure-gradient (ZPG) turbulent boundary layers (TBLs) has been one of the most extensively studied topics in wall turbulence to date (Monkewitz *et al.* 2007; Smits *et al.* 2011; Schlatter & Örlü 2012; McKeon 2017; Jiménez 2018). These efforts have established that the viscous-scaled mean and turbulence statistics of well-behaved smooth-wall ZPG (*i.e.*, canonical) TBLs are a function of the friction Reynolds number, $\mathrm{Re}_\tau$. Here, $\mathrm{Re}_\tau = \delta U_\tau/\nu$, where $\delta$ is the boundary-layer thickness, $\nu$ is the kinematic viscosity, and $U_\tau$ is the mean friction velocity; the latter two quantities are used for viscous scaling of relevant turbulence statistics (denoted by the '+' superscript). While high-$\mathrm{Re}_\tau$ TBLs arise in many practical applications, such as aircraft and ships, they are often subjected to non-canonical conditions, including adverse pressure gradients (APGs). Relative to the canonical case, APGs introduce additional complexity to both the scaling and structure of the TBL (Devenport & Lowe 2022). This forms the focus of the present study, where we identify and characterise the differences between the energetic coherent structures of high-$\mathrm{Re}_\tau$ APG and ZPG TBLs that lead to departures from classical scaling arguments (Smits *et al.* 2011). For the purpose of this study, which is limited to low-to-moderate APG strengths, local streamwise pressure gradients (PGs) are quantified using Clauser's PG parameter, $\beta = (\delta/\rho U_\tau^2)(\mathrm{d}P/\mathrm{d}x)$, where $\delta$ is the displacement thickness, $\rho$ is the density, and $dP/dx$ is the local streamwise PG. Throughout the manuscript, $x$, $y$, and $z$ denote the streamwise, spanwise and wall-normal coordinates, respectively, with $u$, $v$ and $w$ denoting the corresponding velocity fluctuations.

Foundational to this aim, high-$\mathrm{Re}_\tau$ ZPG TBLs comprise a hierarchy of energetic coherent motions spanning small viscous-scaled structures in the near-wall region to larger inertia-dominated motions in the outer region that scale with their distance from the wall and/or the TBL thickness (Smits *et al.* 2011). Throughout this study, 'coherent' motions are identified based on the spatial and/or temporal organisation of velocity fluctuations (such as the recurrent low- and high-streamwise momentum regions induced by strong shear; Adrian *et al.* 2000; Deshpande *et al.* 2023*a*). This hierarchy has typically been characterised using the premultiplied spectrogram ($f\phi^+_{uu}$) of streamwise velocity fluctuations ($u$), an example of which is shown in figure 1(a), plotted as a function of viscous-scaled time scale ($T^+ = U_\tau^2/f\nu$, where $f$ is the frequency) and wall-normal location ($z^+ = zU_\tau/\nu$). This spectrogram corresponds to an APG TBL with $\mathrm{Re}_\tau \approx$ 10,000 reported by Zarei *et al.* (2026*a*). Following Deshpande *et al.* (2023*a*), the spectrogram is divided into four regions (I–IV) to distinguish energy contributions from different types of structures. For a high-$\mathrm{Re}_\tau$ ZPG TBL, region I corresponds to the viscous-scaled near-wall cycle, responsible for intense local production of turbulent kinetic energy in the inner region (Kline *et al.* 1967). Region III corresponds to outer- ($\delta$-) scaled superstructures, or very-large-scale motions (VLSMs), which possess very long time scales and retain a signature at the wall (Hutchins & Marusic 2007; Deshpande *et al.* 2023*a*), becoming statistically significant only at sufficiently high-$\mathrm{Re}_\tau$ ($\gtrsim$ 4000; Smits *et al.* 2011). Region II lies between I and III and is populated by distance-from-the-wall ($z$-) scaled eddies coexisting in the logarithmic (log) region; their time-scale range and energy contributions both increase with $\mathrm{Re}_\tau$. These intermediate-scale motions are geometrically self-similar and have been described using various structural frameworks, including large-scale motions (LSMs; Adrian *et al.* 2000), wall-scaled (*i.e.*, attached) eddies (Baars & Marusic 2020; Yoon *et al.* 2020; Deshpande *et al.* 2021; Yang *et al.* 2026), and uniform-momentum zones (Bross *et al.* 2019; Bautista *et al.* 2019; Gul *et al.* 2020). Since regions II and III are the primary contributors to streamwise turbulent kinetic energy (TKE) and momentum exchange in high-$\mathrm{Re}_\tau$ TBLs (Balakumar & Adrian 2007; Deshpande *et al.* 2023*a*), comparing these regions between high-$\mathrm{Re}_\tau$ APG and ZPG TBLs, forms a central objective of this study.

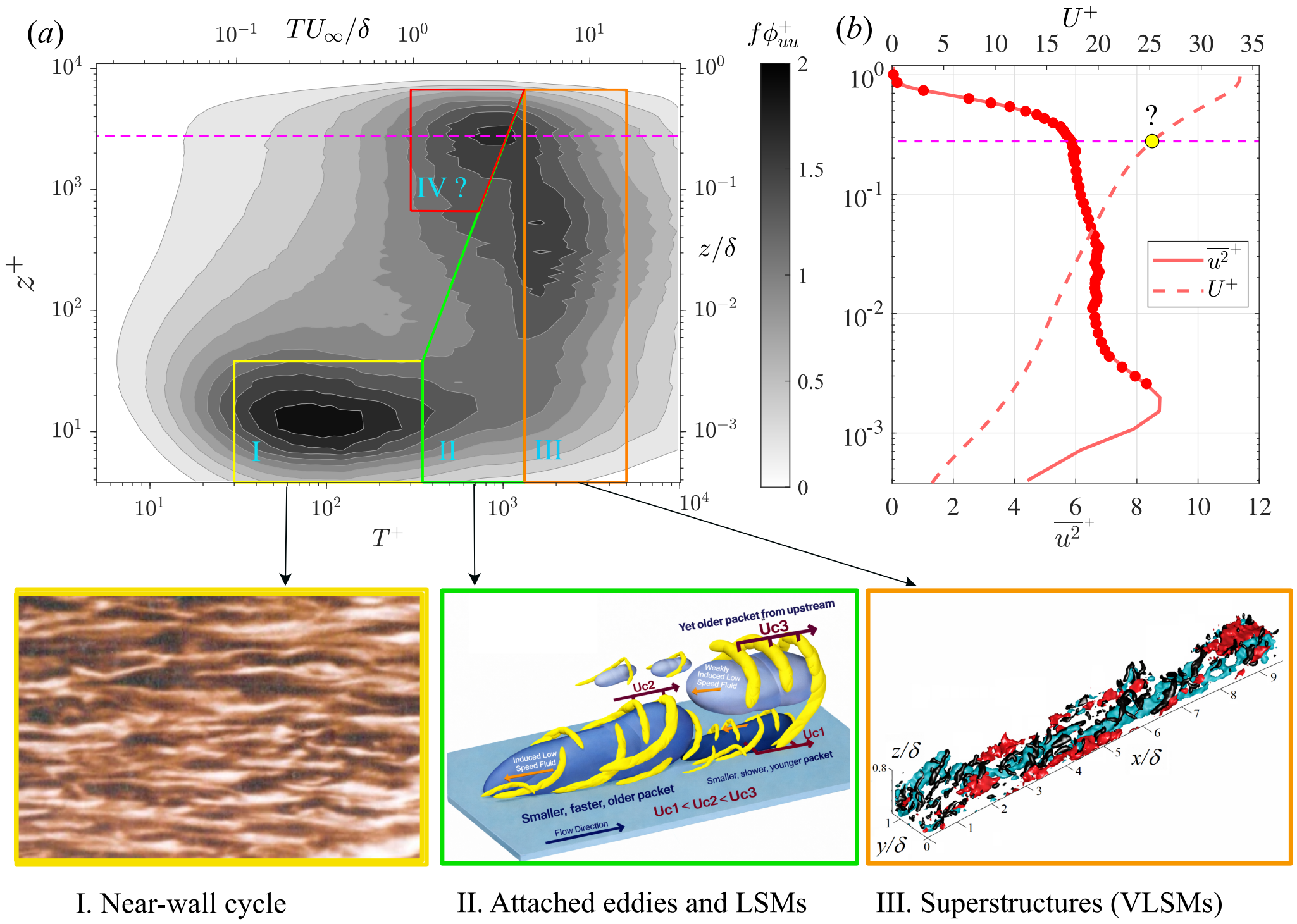

Figure 1: ($a$) Premultiplied spectrogram of streamwise velocity fluctuations ($f\phi^+_{uu}$), as a function of their viscous-scaled time scale ($T^+$) and wall-normal location ($z^+$), for an APG TBL at $Re_\tau \approx 10{,}000$ and $\beta \approx 1.5$ (Zarei *et al.* 2026*a*). Axes are also presented using outer scaling (*i.e.*, based on $\delta$ and $U_\infty$). Coherent motions commonly identified in canonical high-$Re_\tau$ TBLs are indicated for reference: region I corresponds to the near-wall cycle (Kline *et al.* 1967); region II to attached eddies, LSMs, and/or uniform momentum zones (conceptual sketch of Adrian *et al.* 2000); and region III to superstructures/VLSMs (visualised by Dennis & Nickels 2011), following the conceptual framework of Deshpande *et al.* (2023*a*). Region IV denotes APG-energised outer-region motions characterised by scales of $O(\delta)$. ($b$) Profiles of the streamwise mean velocity and variance for the same dataset as ($a$). The dashed magenta line marks the wall-normal location of the spectral peak in the outer region ($z/\delta \approx 0.3$), which closely corresponds to an inflection point in the mean velocity profile (indicated by the yellow circle).

The influence of local APGs on specific scales of motion, and on associated turbulence statistics, is well established in the literature for the case of low-to-moderate $Re_\tau$ TBLs (Devenport & Lowe 2022). For instance, increasing $\beta$ modifies the mean velocity profile (Aubertine & Eaton 2005; Nagano *et al.* 1998; Bobke *et al.* 2017) through strengthening of the wake region, reduction of velocity magnitude within the log region, and decreased mean wall shear stress (until the flow eventually separates). In addition, increasing $\beta$ amplifies both the large- and small-scale motions in the outer region of the TBL (Lee & Sung 2009; Pozuelo *et al.* 2022; Gungor *et al.* 2022). This APG-induced energy amplification of the larger scales results in the emergence of a new spectral peak between $z/\delta = 0.2$–$0.4$ (region IV in figure 1a), distinct from region III (associated with superstructures/VLSMs). This spectral peak also corresponds to the 'outer' peak observed in the Reynolds stress profiles of moderate-to-strong APG TBLs (Pozuelo *et al.* 2022; Gungor *et al.* 2022; Deshpande & Vinuesa 2024). However, distinguishing between regions III and IV is only possible via high-$Re_\tau$ investigations that ensure sufficient scale separation (Deshpande *et al.* 2023*b*; Preskett & Ganapathisubramani 2025), providing further motivation for the flow regime considered here.

A key unresolved question is the extent to which APGs influence the geometrically self-similar scaling of these intermediate-to-large-scale motions (*i.e.*, regions II and III). Alternatively, the differences observed in figure 1(a), relative to a ZPG TBL, may simply arise from superposition of motions associated with the APG-induced energy amplification onto the canonical hierarchy of inertia-dominated motions (Deshpande *et al.* 2021, 2023*a*). One established approach for investigating this is through two-point cross-correlation-based analyses, such as the linear coherence spectrum (LCS; Bendat & Piersol 2011). Here, we deploy the LCS framework, similar to that implemented by Baars & Marusic (2020), to quantify the time-scale-resolved spectral energy that is stochastically coherent, or incoherent, between two wall-normal locations. Specifically, a fixed reference point is chosen within the near-wall region ($z^+ < 7$) to construct a time-scale-dependent filter that decomposes $\phi_{uu}(z)$ into stochastically wall-coherent and wall-incoherent components. As will be shown ahead, this approach provides a direct means of distinguishing energy contributions associated with APG-induced energy amplification from the underlying hierarchy of wall-coherent motions common to both ZPG and APG TBLs (*i.e.*, regions II and III). The framework also enables examination of whether APGs alter the established geometric self-similarity of these wall-coherent motions reported by Baars & Marusic (2020) and Deshpande *et al.* (2021).

While regions I, II and III are associated with distinct coherent structures (figure 1a), an open question is which structure(s) is responsible for the enhanced TKE in region IV of low-to-moderate APG TBLs? Studies of low-$\mathrm{Re}_\tau$ TBLs subjected to strong $\beta$ ($\gtrsim$ 6) have reported a possible flow-instability mechanism associated with an inflection point in the mean velocity profile (Bradshaw 1967; Elsberry *et al.* 2000; Gungor *et al.* 2016; Kitsios *et al.* 2017; Schatzman & Thomas 2017; Silva & Wolf 2024). This instability has commonly been linked to compact, spanwise-oriented, clockwise-rotating vortical structures forming so-called *embedded shear layers* in the outer region of the TBL. This interpretation also aligns naturally with the composite mean velocity profile of Coles (1956), which represents the mean flow as a combination of a viscous near-wall region and a wall-independent wake region. Although the experimental findings of Schatzman & Thomas (2017) were limited to relatively strong APGs, they hypothesised that embedded-shear-layer-type structures may also coexist with other coherent structures in low-to-moderate APG TBLs. However, the mere existence of an inflection point in the mean profile cannot confirm the presence/absence of distinct shear-layer-type motions in the flow field, a limitation considered in the present study. Here, we characterise the outer-region motions associated with the APG-induced energy amplification to directly assess whether they are spatially and temporally organised in a manner consistent with an embedded shear layer. If so, the enhanced wake component of the composite mean velocity profile may be physically interpreted as the manifestation of these APG-amplified outer-region structures, providing a basis for future modelling efforts. It is noteworthy, however, that shear layers and locally inflectional velocity profiles have also been identified in conditionally averaged ZPG TBL flow fields, particularly at interfaces between uniform momentum zones in the outer region (Adrian *et al.* 2000; Bautista *et al.* 2019; Gul *et al.* 2020). Accordingly, the present study does not distinguish whether the motions associated with the APG-induced energy amplification correspond to a *new* shear-layer structure or simply to amplification of existing shear-layer structures associated with uniform momentum zones.

Besides practical relevance, investigating APG TBLs at high-$\mathrm{Re}_\tau$ is important because Reynolds number and APG strength exert competing influences on TKE (Vinuesa *et al.* 2018; Pozuelo *et al.* 2022; Deshpande *et al.* 2023*b*). At fixed $\beta$, increasing $\mathrm{Re}_\tau$ energises large-scale motions in the outer region while reducing the relative energy of coexisting small-scale motions (Deshpande *et al.* 2023*b*). By contrast, at fixed $\mathrm{Re}_\tau$, increasing $\beta$ enhances both small- and large-scale motions in the outer region. However, most previous studies of APG TBL flow structure have been conducted either at low $\mathrm{Re}_\tau$ (Nagano *et al.* 1998; Volino *et al.* 2007; Yoon *et al.* 2020; Pozuelo *et al.* 2022), where scale separation between regions I and IV is limited, and/or under conditions with strong or unique PG history effects (Bross *et al.* 2019; Tanarro *et al.* 2020; Knopp *et al.* 2021; Romero *et al.* 2022; Parthasarathy & Saxton-Fox 2025). For example, Tanarro *et al.* (2020) and Gungor *et al.* (2022) performed LCS analyses on APG TBLs at low-$\mathrm{Re}_\tau$ and high-$\beta$, identifying that motions associated with the APG-induced energy amplification are predominantly wall-incoherent and reside in the outer region; however, they could not definitively comment on the scaling of wall-coherent motions (*i.e.*, regions II and III, which are not clearly separated at low $\mathrm{Re}_\tau$). Owing to practical limitations in experimental and numerical setups, isolating and generalising the effect of local APG independent of PG history remains challenging, despite both influencing the outer-region TKE (Bobke *et al.* 2017; Preskett & Ganapathisubramani 2025). In the present study, we overcome this challenge by leveraging a unique experimental setup that minimises upstream PG history effects (Deshpande *et al.* 2023*b*), enabling systematic examination of local APG effects on inertia-dominated motions at high $\mathrm{Re}_\tau$.

### 1.1. *Present contributions*

The present study identifies and characterises the flow structures and energy contributions associated with APG-induced energy amplification in the outer region of high-$\mathrm{Re}_\tau$ TBLs using complementary experimental datasets acquired via hot-wire (HW) measurements and particle image velocimetry (PIV). The selected flow conditions ensure clear separation between inner and outer scales, permitting unambiguous assessment of how local APGs modify the canonical hierarchy of coherent structures across regions II and III, independent of PG history effects. First, the influence of APGs on the hierarchy of inertia-dominated motions is examined using two-point HW measurements and the LCS. This approach quantifies the correlation, if any, between outer-scaled motions associated with the APG-induced energy amplification (region IV) and other structures coexisting in high-$\mathrm{Re}_\tau$ TBLs. Second, we test whether the TKE added upon imposition of a moderate APG is consistent with the embedded shear layer hypothesis by conditionally averaging spatially resolved velocity fields from a previously published PIV dataset (Lindić *et al.* 2025). Conditional averaging is performed using intense vortical events identified via the swirling-strength criterion, following related approaches (Wu & Christensen 2006; Dennis & Nickels 2011; Lee 2017). This analysis also directly links the inflection point in the mean velocity profile to the flow structures associated with the outer energetic peak in the Reynolds stress profile and premultiplied energy spectra (at least under moderate APG conditions; figure 1b).

## 2. Experimental setup, methodology and boundary layer characterisation

All experiments were conducted in the high-Reynolds-number boundary layer wind tunnel at the University of Melbourne (Marusic *et al.* 2015; Deshpande *et al.* 2023*b*). The test-section ceiling is equipped with an adjustable array of air-bleed slots spanning the full width and distributed uniformly in the streamwise direction. A nominal ZPG TBL is obtained using a single high-porosity mesh at the outlet with all air-bleed slots fully open (Marusic *et al.* 2015). Following Deshpande *et al.* (2023*b*), the test-section static pressure is controlled by varying the outlet blockage. Specifically, different APG conditions are generated by adding either one or three low-porosity screens at the outlet, in addition to the high-porosity mesh. These configurations are referred to as APG1 (green symbols in figure 2a) and APG3 (red symbols in figure 2a), respectively. As shown by the streamwise pressure coefficient ($C_P$) profiles in figure 2(a), both APG cases exhibit nominal ZPG conditions upstream of the APG development region ($x < 9$ m), ensuring that the APG is introduced under established high-Reynolds number conditions with minimal PG history effects (Deshpande *et al.* 2023*b*).

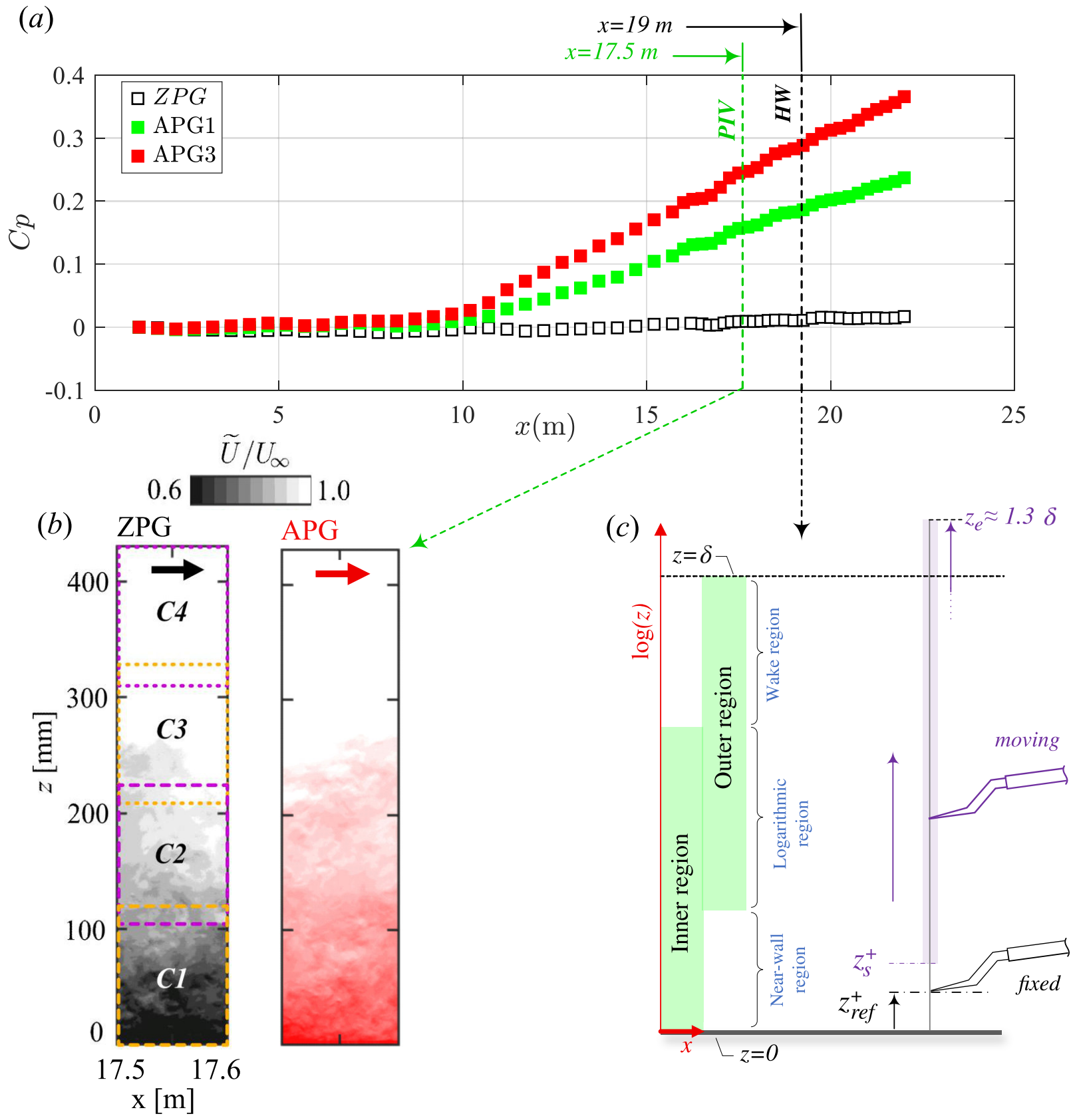

Figure 2: ($a$) Measured $C_P(x)$ profiles for the ZPG, APG1, and APG3 cases. The vertical green dashed line marks the location of the PIV measurement, while the black dashed line indicates the location of the HW measurements. ($b$) Field of view of the PIV setup, adapted from Lindić *et al.* (2025), showing instantaneous streamwise velocity snapshots for the ZPG (grey shading) and APG3 (red shading) cases across the full TBL, obtained by stitching fields across four cameras (C1–C4). ($c$) Schematic illustrating the experimental setup for the two-point synchronised hot-wire measurements. Flow conditions are documented in table 1.

Here, the pressure coefficient is defined as $C_P(x) = 1 - U_\infty^2(x)/U_\infty^2(0)$, where $U_\infty(x)$ is the mean free-stream velocity at $x$, with $x = 0$ denoting the test-section inlet. This study reports new two-point HW measurements (§2.1) and previously published 2-D snapshot PIV measurements (§2.2; Lindić *et al.* 2025) of high-$Re_\tau$ ZPG and APG TBLs, conducted at $x$ $\approx$ 19:m and 17.5:m, respectively (figure 2a).

### 2.1. *Two-point hot-wire measurements*

Two-point measurements were performed for the ZPG, APG1 and APG3 cases (*i.e.*, with different local values of $\beta$) at matched friction Reynolds numbers of $\mathrm{Re}_\tau \approx$ 10,000, as documented in table 1. The experimental setup was directly inspired by a previous study from the research group (Baars & Marusic 2020), comprising two vertically (wall-normal) offset hot-wire probes with negligible spanwise and streamwise separation (figure 2$c$). Following Baars & Marusic (2020), a near-wall reference probe was fixed at $z^+_{\mathrm{ref}} \lesssim 7$ (nominally within the viscous sublayer), while a second probe (referred to as the traversing probe) was moved from $z_s^+ \approx 30$ to $z_e/\delta = 1.3$ across several logarithmically spaced wall-normal locations to quantify coherence between motions in the near-wall region and the remainder

**Turbulent boundary layer information: two-point hot-wire measurements (x = 19 m)**

| PG cases | $Re_\tau$ | $\beta$ | $U_\tau$ (m/s) | $\delta$ (mm) | $\delta^*$ (mm) | $z^+_{ref}$ | $z^+_s - z^+_e$ | $\Delta T^+$ | $l^+$ | Plotting style |
|---|---|---|---|---|---|---|---|---|---|---|
| ZPG | 10600 | 0 | 0.436 | 360 | 38.1 | 6.5 | $32 - 1.3\,\delta^+$ | 0.25 | 14.5 | ■ — |
| APG1 | 10300 | 0.7 | 0.374 | 413 | 53.4 | 5.4 | $29 - 1.2\,\delta^+$ | 0.18 | 12.5 | ♦ — |
| APG3 | 10500 | 1.6 | 0.337 | 460 | 65.8 | 3.5 | $27 - 1.2\,\delta^+$ | 0.15 | 11.5 | ● — |

**Turbulent boundary layer information: 2-D snapshot PIV (x = 17.5 m)**

| PG cases | $Re_\tau$ | $\beta$ | $U_\tau$ (m/s) | $\delta$ (mm) | $\delta^*$ (mm) | Field of view $(x \times z)$ (mm)$^2$ | Sampling window $(\Delta x^+ \times \Delta z^+)$ | Plotting style |
|---|---|---|---|---|---|---|---|---|
| ZPG | 9200 | 0 | 0.40 | 343 | 43.4 | $104 \times 440$ | $18 \times 18$ | — |
| APG3 | 8700 | 1.45 | 0.33 | 392 | 65.4 | $104 \times 440$ | $11 \times 11$ | — |

Table 1: Characteristics of the turbulent boundary layers investigated in the present study. PIV sampling windows corresponds to a fixed interrogation window size of $24 \times 24$ pixels ($0.53 \times 0.53$ mm$^2$) with 50% overlap. Refer to Lindić *et al.* (2025) for more details.

of the TBL. Here, $z^+_s$ and $z^+_e$ denote the start and end locations of the traversing probe, respectively. Particular attention was given to ensuring adequate wall-normal resolution near $z/\delta \approx 0.3$–$0.4$, where local APG effects are known to produce the strongest APG-induced energy amplification of turbulent motions (Deshpande *et al.* 2023*b*) (figure 1a). The smallest wall-normal gap between the reference probe and the starting location of the traversing probe ($z^+_s \approx 30$) was sufficient to avoid probe-interference and blockage effects, consistent with the assessment of Baars & Marusic (2020). For completeness, in addition to the two-point measurements, a set of single-point hot-wire measurements was also acquired for matched $\mathrm{Re}_\tau$ and PG conditions across $5 \lesssim z^+ \lesssim \delta^+$ to validate the two-point measurements using conventional TBL statistics. These single-point measurements were also used to estimate the boundary-layer thickness ($\delta$) using the methodology proposed by Lozier *et al.* (2025) based on streamwise velocity skewness (see table 1). Oil-film interferometry was used to provide an accurate and independent estimate of $U_\tau$, further details of which are documented in previous studies from the group (Lindić *et al.* 2025; Zarei *et al.* 2026*a*).

The hot-wire probes were fabricated in-house using 2.5 $\mu$m diameter Wollaston wires mounted on standard Dantec 55P15 probes. The wires were etched to a nominal sensing length of $l = 0.5$ mm, yielding a sensor aspect ratio of $l/d = 200$ and a viscous-scaled sensing length of $l^+ = 11.5$–$14.5$ across all cases. For the present low-to-moderate APG conditions at sufficiently high Reynolds number ($\mathrm{Re}_\tau \approx 10{,}000$), the influence of APG on small-scale energy in the outer region is weak compared with that observed at lower $\mathrm{Re}_\tau$ (Deshpande *et al.* 2023*b*). The primary APG-induced energy amplification is instead concentrated in large-scale outer-region motions, which are fully resolved by the present hot-wire measurements. The hot-wire signals were sampled for a total duration, $T_S$, satisfying $T_S U_\infty/\delta \geqslant 20{,}000$, to ensure statistical convergence of the energy spectra representing the hierarchy of large- and very-large-scale motions (Deshpande *et al.* 2023*b*). Signals were sampled at 50 kHz and low-pass filtered at 25 kHz, resulting in viscous-scaled sampling intervals $\Delta T^+ = U^2_\tau/(f_S \nu)$ in the range 0.15–0.25, sufficient to resolve the small-scale turbulence. Further details of the hot-wire methodology and calibration procedures are provided in Deshpande *et al.* (2023*b*).

At the same streamwise location ($x = 19$ m), complementary two-point hot-wire measurements for the present ZPG and APG3 configurations were also reported by Lozier *et al.* (2026*b*). In that arrangement, the reference probe was fixed in the outer region ($z/\delta \approx 0.4$), while the traversing probe moved from the near-wall region ($z^+ \approx 5$) to the boundary-layer edge. These measurements enabled characterisation of the linear coupling between outer-region motions and the rest of the TBL through the LCS. Their results indicated that motions coherent with the outer-region reference location were predominantly confined to the outer/wake region ($0.2 \lesssim z/\delta \lesssim 0.6$) and did not leave a measurable footprint at the wall across any time scale. Although these measurements were useful for identifying motions locally coherent within the wake, it was found that a single outer-region reference location was insufficient to capture all of the additional streamwise TKE associated with APG-induced energy amplification in the outer region. For this reason, the present study focuses on the LCS using a near-wall reference location, which we find sufficient to segregate the energy associated with wall-incoherent motions (comprising most of the APG-induced energy amplification) from the wall-coherent motions (refer to §3 for further details).

### 2.2. *Two-dimensional snapshot particle image velocimetry*

The 2-D PIV datasets comprise spatially well-resolved velocity fields spanning the entire boundary-layer thickness for both the ZPG and APG3 cases (figure 2b), thereby complementing the temporally resolved hot-wire dataset. Although the streamwise measurement location for the PIV experiment differs from that of the HW experiment owing to practical limitations, the measurement conditions for both ZPG and APG3 were quasi-matched between the two datasets (table 1). The PIV system employed four vertically offset Imager CX-25 cameras (see dashed squares in figure 2b), achieving a digital resolution of 22 $\mu$m/pixel. This arrangement provided complete coverage of the TBL thickness; *i.e.*, the stitched field of view ($x \times z \sim 104 \times 440$ mm$^2$) spans the full $\delta$ for both cases. The corresponding viscous-scaled spatial resolution, reported in table 1, is $(\Delta x^+ \times \Delta z^+) \approx (18 \times 18)$ for the ZPG case and $(11 \times 11)$ for the APG3 case. This resolution is comparable to, or finer than, that used in previous experimental studies aimed at identifying vortical structures in wall-bounded turbulence (Wu & Christensen 2006), and is therefore sufficient to resolve intense vorticity regions and vortex cores in the present datasets. Representative instantaneous streamwise velocity fields for the ZPG and APG3 cases are shown in figure 2(b). Both PIV datasets were validated by comparison of one-point statistics with published multi-wire anemometry measurements in Lindić *et al.* (2025), to which the reader is referred for additional technical details of the PIV setup. The key TBL parameters obtained from both the hot-wire and PIV measurements are summarised in table 1.

### 2.3. *Characterisation of adverse-pressure-gradient turbulent boundary layer*

With the HW and PIV setups established, we now present conventional one-point statistics to contextualise the two-point analyses discussed later in §3 and §4. Figure 3 depicts the time- and ensemble-averaged turbulence statistics obtained from the HW (figures 3a,b) and PIV (figures 3c–f) measurements, respectively, all of which broadly follow expected trends when comparing ZPG and APG TBLs (Deshpande *et al.* 2023*b*; Lindić *et al.* 2025). Figure 3(a) shows the deviation of the mean velocity from the classical log law, $U^+ - U^+_{\log}$, highlighting the influence of APG on the mean velocity profile. Here, the classical log law is given by $U^+_{\log} = (1/\kappa)\ln(z^+) + B$, where $\kappa$ and $B$ are the von Kármán and additive coefficients, respectively, taken as 0.39 and 4.3 for the ZPG case (Zarei *et al.* 2026*a*). Compared with the ZPG case, increasing APG leads to a stronger wake, earlier departure from logarithmic scaling, and increasing negative deviation within the log region. Despite this shift, the profile slopes in the log region remain nominally matched to the ZPG case (within experimental uncertainty), indicating that $\kappa$ likely remains unchanged and that the shift instead originates from a reduction in the additive coefficient, $B$. The upper bound of the log region is identified using the threshold criterion $|U^+ - U^+_{\log}| \geqslant 0.05$, yielding a ZPG log-region extent consistent with previous literature ($z^+ \approx 0.15\,\mathrm{Re}_\tau$; Marusic *et al.* 2015; Deshpande *et al.* 2023*b*). Under APG conditions, however, the mean velocity departs from the classical log law at lower wall-normal locations (*i.e.*, closer to the wall; inset of figure 3a, Devenport & Lowe 2022).

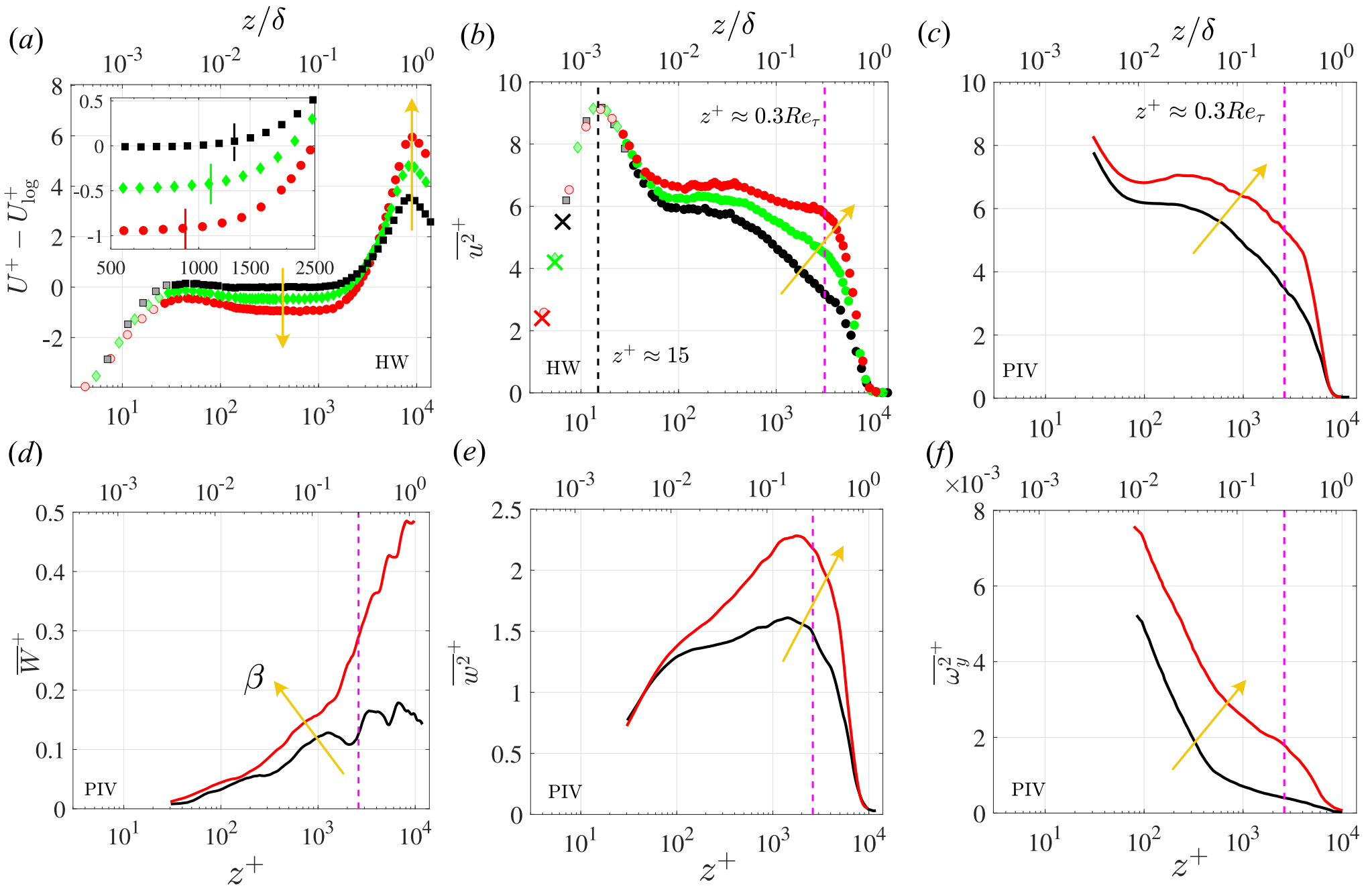

Figure 3: (*a*) Deviation of the mean streamwise velocity from the log-law, $U^+ - U^+_{\log}$, and (*b*) corresponding streamwise velocity variance, $\overline{u^2}^+$, as a function of $z^+$ obtained from HW measurements. Vertical solid lines in the inset of (*a*) mark the departure of the mean velocity profile from the log law. Cross symbols indicate the fixed reference HW locations ($z^+_{\text{ref}}$) while the pale-coloured symbols denote the additional wall-normal locations sampled by the traversing wire to obtain the complete profile. Profiles of (*c*) $\overline{u^2}^+$, (*d*) the mean wall-normal velocity $\overline{W}^+$, (*e*) its variance $\overline{w^2}^+$, and (*f*) spanwise vorticity variance $\overline{\omega_y^2}^+$ obtained from PIV. Dashed magenta lines denote the outer-region location, $z/\delta \approx 0.3$, associated with APG-energised motions (see figure 1). The black, green, and red curves correspond to the ZPG, APG1, and APG3 cases, respectively.

Figures 3(b,c) compare the $\overline{u^2}^+$ profiles obtained from the HW and PIV measurements, respectively. In the near-wall region, the $\overline{u^2}^+$ for the ZPG and APG cases nearly collapse up to $z^+ \approx 30$ (figure 3*b*), indicating that APG effects are negligible there. This also justifies the choice of the starting traversing-probe location at $z^+_s \approx 30$. In contrast, pronounced differences emerge in the outer region, particularly around $z/\delta \approx 0.3$ (indicated by the magenta dashed line). This reflects the presence of APG-amplified outer-region motions, either through the emergence of new structures or through modification of structures already present under ZPG conditions, or both. Consistent with this interpretation, figures 3(d–f) show that the mean wall-normal velocity ($W^+$), the wall-normal velocity variance ($\overline{w^2}^+$), and the spanwise vorticity variance ($\overline{\omega_y^2}^+$) all increase in the outer region under APG conditions, consistent with previous studies (Romero *et al.* 2022; Wei *et al.* 2023). Here, we define the spanwise vorticity as $\omega_y = \partial u/\partial z - \partial w/\partial x$.

The elevated vorticity fluctuations suggest that APG increases the number and/or intensity of outer-region vortical structures, which are closely linked to the production of Reynolds shear stress (Lee & Sung 2009) and other shear-layer-related effects (Adrian *et al.* 2000; Schatzman & Thomas 2017; Gul *et al.* 2020; Silva & Wolf 2024). These observations are also consistent with the results of Deshpande & Vinuesa (2024), who found motions associated with APG-induced energy amplification to be particularly enhanced near the $\overline{w^2}^+$ peak, with intensified sweep and ejection events playing a central role in energy transfer. Motivated by these findings, we use the PIV dataset to further analyse $u$- and $w$-fluctuations (and the corresponding sweep and ejection events) in the $x$–$z$ plane to assess the applicability of the embedded shear layer hypothesis.

## 3. Effect of adverse pressure gradients on the wall-scaled eddy hierarchy

This section uses the LCS to address the following question: do the APG-energised motions in the outer region (region IV in figure 1a) modify/influence the existing hierarchy of structures established for canonical high-$Re_\tau$ TBLs (such as LSMs and superstructures)? Section §3.1 first briefly outlines the LCS framework and its implementation in the present study. Specifically, when the reference probe is fixed close to the wall, the LCS can be used to linearly decompose the streamwise energy spectra across the TBL into its stochastically wall-coherent and incoherent components on a scale-by-scale basis (Baars & Marusic 2020). Based on the LCS, §3.2 then examines the wall-coherent motions in region II of the APG TBL (figure 1a) focusing on their geometric self-similarity, and whether any changes can be explained through alteration of the mean inclination of the structures. Finally, §3.3 uses the spectral decomposition of wall-coherent and incoherent energy to assess the influence of APG on the very-large-scale motions (region III) and the emergence of the new region IV.

### 3.1. *The linear coherence spectrum*

To quantify the linear coupling between turbulent motions at a given wall-normal location with the near-wall region, we compute the LCS, denoted by $\gamma_L^2(z^+, z_{ref}^+; T^+)$, using the two-point hot-wire measurements of synchronously acquired velocity time series, $u(z)$ and $u(z_{ref})$. The LCS is calculated as (Bendat & Piersol 2011):

$$\gamma_L^2(z, z_{ref}; T) \equiv \frac{\left|\{\hat{U}(z;\, T)\, \hat{U}^*(z_{ref};\, T)\}\right|^2}{\{|\hat{U}(z;\, T)|^2\}\ \{|\hat{U}(z_{ref};\, T)|^2\}}, \tag{3.1}$$

where $\hat{U}$ is the Fourier transform of $u(z)$, the asterisk $*$ denotes the complex conjugate, $\{\cdot\}$ indicates ensemble averaging, and $|\cdot|$ represents the modulus. By definition, the normalised coherence is limited to $0 \leqslant \gamma_L^2 \leqslant 1$. In terms of physical interpretation, $\gamma_L^2$ reflects the fraction of the spectral energy at a given $z^+$ and $T^+$ that is linearly coherent with the near-wall reference signal. For instance, $\gamma_L^2 = 0$ indicates that velocity fluctuations at a given scale exhibit no consistent phase relationship, and therefore have no stochastic correlation between $z$ and $z_{ref}$.

### 3.2. *Geometric self-similarity and inclination of wall-coherent motions*

The existence of a geometrically self-similar hierarchy of wall-coherent motions (region II in figure 1) requires the linear coherence spectrum in the log region to collapse when expressed in $z$-scaled coordinates, *i.e.*, $\gamma^2_L$ should depend solely on the ratio of length scales, $TU(z)/z$. As shown by Baars & Marusic (2020) for the ZPG TBL, the coherence magnitude in the log region is well described by an empirically obtained log-based expression, with its zero-coherence intercept defining the characteristic aspect ratio ($AR$) of the wall-coherent motions in the streamwise–wall-normal plane:

$$\gamma^2_L = C_1 \ln\left(\frac{TU(z)}{z}\right) + C_2, \qquad \text{with} \qquad AR \equiv \left.\frac{TU(z)}{z}\right|_{\gamma^2_L=0} = \exp\left(\frac{-C_2}{C_1}\right), \tag{3.2}$$

where $C_1$ and $C_2$ are constants and the streamwise length scale ($l_x$) is estimated by assuming Taylor's hypothesis ($l_x \sim TU(z)$). As shown in figure 4(a), the present $\gamma^2_L$ curves exhibit an apparent collapse when plotted against $TU(z)/z$ for both the ZPG and APG cases. This indicates that the wall-coherent motions in the log region retain an approximately geometrically self-similar organisation even under moderate APG conditions. The associated $AR$, however, decreases systematically with increasing $\beta$, from $AR \approx 14$ for ZPG, consistent with Baars & Marusic (2020), to $AR \approx 11.5$ for APG3. This decrease implies a reduction in the characteristic streamwise extent relative to the wall-normal extent, suggesting that the representative wall-coherent structures become relatively more inclined with the wall under APG conditions (figure 4b). This interpretation is also consistent with the enhanced wall-normal velocity ($W^+$) noted for an APG TBL (figure 3d), which will increase the vertical momentum transfer and make the structures appear more inclined (Vinuesa *et al.* 2018).

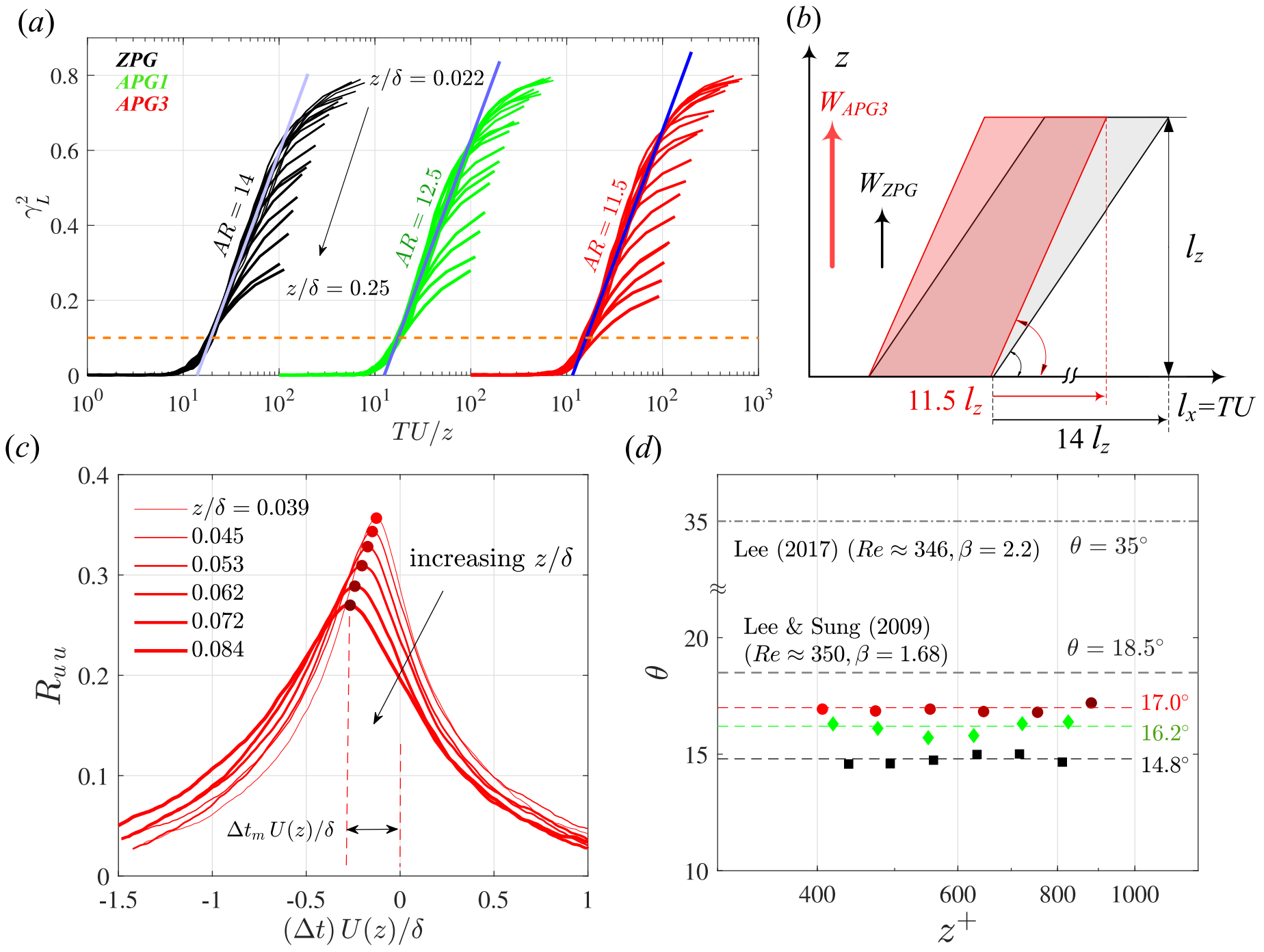

Figure 4: (*a*) Linear coherence spectrum $\gamma^2_L$ as a function of $TU(z)/z$ for the ZPG (black), APG1 (green), and APG3 (red) cases, evaluated over wall-normal positions corresponding to $z/\delta \approx 0.02$–$0.25$. Solid blue lines indicate the best logarithmic fits (varying intensity blue indicates different slopes), while the horizontal dashed orange line denotes $\gamma^2_L = 0.1$, which is used as a low-coherence threshold to identify deviations from self-similar scaling. (*b*) Schematic illustration of the streamwise-to-wall-normal geometry of representative wall-coherent structures (in the log-region) for the ZPG and APG3 cases. (*c*) Cross-correlations $R_{uu}$ between $u(z)$ and $u(z_{\rm ref})$ for locations within the log region, with peak values indicating the temporal offset, $\Delta t_m U(z)/\delta$. (*d*) Mean inclination angle of wall-coherent structures in the log region for the present ZPG and APG cases, together with estimates from the low-Re$_\tau$ data in the literature.

To directly quantify the mean angle of wall-coherent structures in the log-region, we obtain cross-correlations ($R_{uu}$) of the streamwise velocity fluctuations at ($z$) and $z_{\text{ref}}$ following (Brown & Thomas 1977; Marusic & Heuer 2007):

$$R_{uu}(z_{\text{ref}}, z; \Delta t) = \frac{\{u(z_{\text{ref}}; t)\, u(z; t + \Delta t)\}}{\sqrt{\{u^2(z_{\text{ref}})\}}\sqrt{\{u^2(z)\}}}, \tag{3.3}$$

where the $\Delta t$ is the temporal delay and the mean inclination angle ($\theta$) is calculated as:

$$\theta = \arctan\left(\frac{z - z_{\text{ref}}}{\Delta t_m U(z)}\right). \tag{3.4}$$

Here, $\Delta t_{\text{m}} U(z)$ is an equivalent streamwise length scale of the maxima of the correlation by assuming Taylor's hypothesis (see Marusic & Heuer 2007). $R_{uu}$ profiles for the APG3 case at representative wall-normal locations in the log region are shown in figure 4(c), with the maxima/peaks marked by red circles. The systematic shift in time of the peak (-$\Delta t_m$) with increasing $z/\delta$ is consistent with Marusic & Heuer (2007). The corresponding mean inclination angles for all $\beta$ cases have been presented in figure 4(d), indicating no discernible/systematic variation with $z$. For the ZPG case, the obtained value ($\theta \approx 14.8°$) agrees well with previous findings, with a relatively small scatter (Marusic & Heuer 2007). Consistent with past findings at low-$\text{Re}_\tau$ (Lee 2017; Lee & Sung 2009), $\theta$ systematically increases with an increase in $\beta$, corresponding to the observed reduction in $AR$. However, the present high-$\text{Re}_\tau$ dataset provides new evidence of a monotonic increase in $\theta$ with increasing local $\beta$, confirming its independence from upstream history effects.

It is noteworthy that, although the collapse of $\gamma_L^2$ persists even in the presence of low-to-moderate APG, the range of wall-normal locations exhibiting self-similarity tends to reduce with increasing $\beta$. Indeed, on comparing $\gamma_L^2$ within the same outer-normalised wall-normal range ($z/\delta \approx 0.02$–$0.25$) in figure 4(a), the APG cases can be found to depart from the geometrically self-similar scaling at a lower $z/\delta$, while also reaching the practical low-coherence threshold ($\gamma_L^2 \approx 0.1$) at a smaller $z/\delta$ than the ZPG case. This indicates that although the imposition of a moderate APG does not nullify the geometric self-similarity of the wall-coherent motions, the APG-energised outer-region dynamics do limit the wall-normal extent over which the self-similarity is observed, through their growing penetration into the logarithmic region with increasing $\beta$. This is also consistent with the departure of the mean velocity profile from the classical log law in figure 3(a), which was noted to occur at gradually lower wall-normal locations with increasing $\beta$.

### 3.3. *Spectral decomposition of wall-coherent and wall-incoherent energy*

With the intermediate-scales continuing to exhibit geometric self-similarity, we next investigate whether the additional APG-induced energisation in the outer-region is limited solely to wall-incoherent motions, or whether it also influences the magnitude of the wall-coherent hierarchy. To examine this, the streamwise energy spectra are decomposed into their wall-coherent and incoherent components using the LCS-based filter (Tinney *et al.* 2006):

$$\phi^+_{uu}(z^+; T^+) = \underbrace{(\gamma_L^2)\, \phi^+_{uu}(z^+; T^+)}_{\text{wall-coherent}} + \underbrace{(1 - \gamma_L^2)\, \phi^+_{uu}(z^+; T^+)}_{\text{wall-incoherent}}. \tag{3.5}$$

Figures 5(a–c) show the premultiplied energy spectra $f\phi^+_{uu}$ together with the $\gamma_L^2$ contours for the ZPG, APG1, and APG3 cases, while figures 5(d–f) present contours of the corresponding

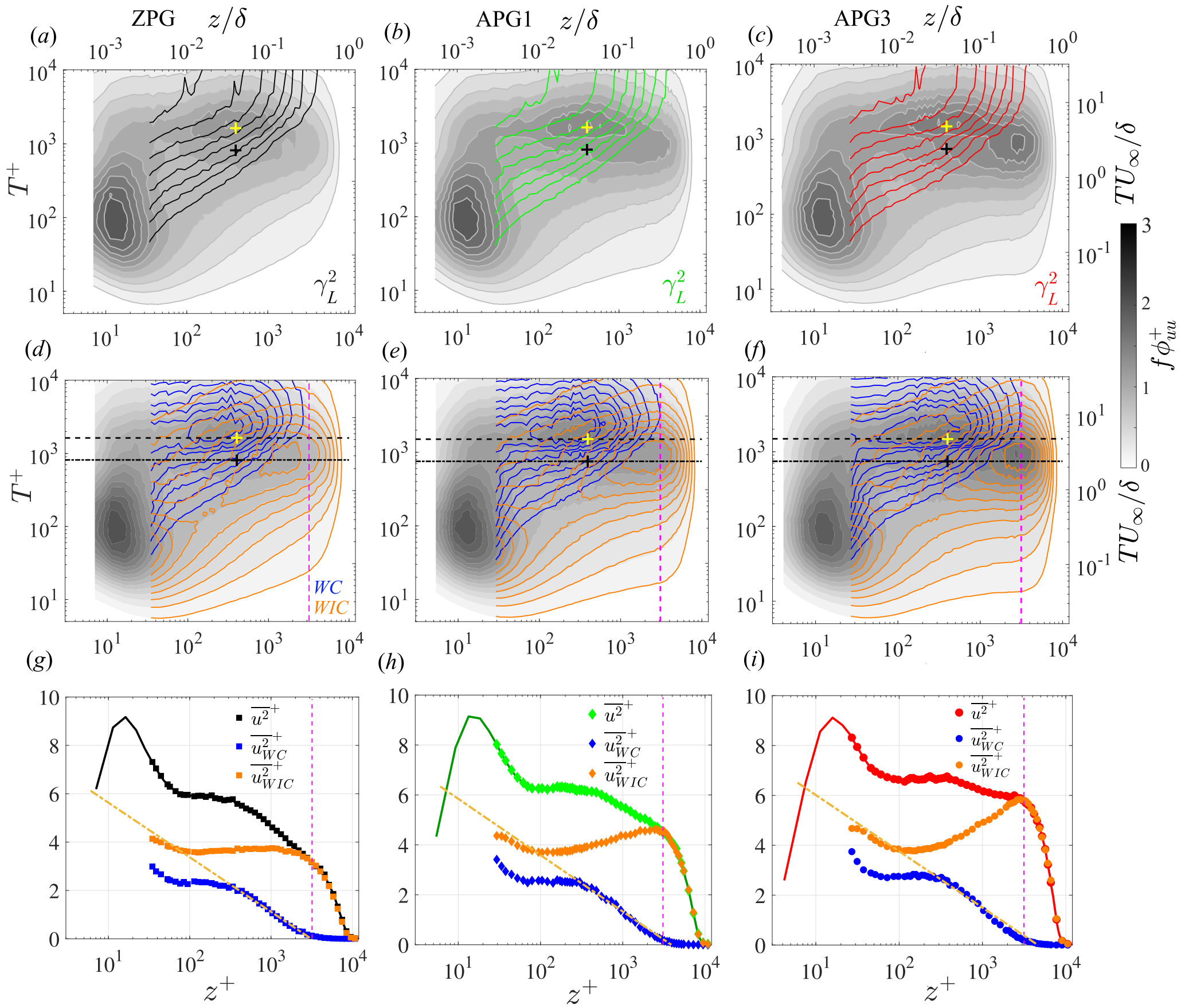

Figure 5: Premultiplied energy spectra, $f\phi^+_{uu}$ (a-f; in grey-scaling), together with the $\gamma^2_L$ contours for the ZPG, APG1, and APG3 cases (a-c) in respective colours. Corresponding wall-coherent and incoherent contributions to (d–f) $f\phi^+_{uu}$ and (g–i) $\overline{u^2}^+$. In (a–c), the black (+) and yellow (+) plus symbols respectively mark the intersections of the middle of the log region, $z^+_m = 3.9\mathrm{Re}^{0.5}_\tau$, with the characteristic time scales associated with APG-induced energy ($T^+/(U^2_\tau/U_\infty\nu) = 2.4\delta$; Deshpande *et al.* 2023*b*) and superstructures ($T^+/(U^2_\tau/U_\infty\nu) = 4.8\delta$; Marusic *et al.* 2015). The same time scales are respectively indicated in (d–f) using horizontal dashed and dash-dotted lines, while the vertical magenta line marks $z/\delta \approx 0.3$. In (d-f), the wall-coherent contributions are shown by blue contours plotted at identical levels for all three cases (0.1:0.1:0.9), while the wall-incoherent contributions are plotted using the same contour levels as $f\phi^+_{uu}$. The dash-dotted golden lines in (g–i) depict the logarithmic decay of $\overline{u^2}^+ = B_1 - A_1 \ln(z/\delta)$ with $A_1 = 0.98$ (Baars & Marusic 2020; Deshpande *et al.* 2021).

wall-coherent and incoherent contributions to $f\phi^+_{uu}$. In these figures, the black and yellow '+' symbols respectively represent the time scales associated previously with APG-induced energy ($TU_\infty \approx 2.4\delta \approx 3\delta_{99}$) and very-large-scale wall-coherent motions/superstructures ($TU_\infty \approx 4.8\delta \approx 6\delta_{99}$) at a well-accepted reference location in the log region, $z^+ = 3.9\mathrm{Re}^{0.5}_\tau$ (Hutchins & Marusic 2007). These points are used as a common reference by which to compare the energy levels of the wall-coherent motions in the log region relative to the APG-energised outer-scaled contributions. In figures 5(d–f), the same time scales are also indicated by the horizontal dashed ($TU_\infty = 4.8\delta$) and dash-dotted ($TU_\infty = 2.4\delta$) lines, while the vertical magenta line marks the nominal location of the maximum APG-energisation ($z/\delta \approx 0.3$). Notably, the wall-coherent contribution, shown by the blue contours in figures 5(d–f), is plotted with the same contour levels for all three cases, thereby permitting direct comparison of their energy level and spatial organisation across the ZPG, APG1, and APG3 cases.

The pre-established hierarchy of coherent motions (as defined in figure 1a) remains clearly identifiable across all three cases in figures 5(a–c), comprising the near-wall cycle, the attached eddies/LSMs, and superstructures/VLSMs. Within the log region, where the geometrical self-similarity of wall-coherent motions is expected to be most evident, the coherence level associated with the two reference time scales (black and yellow '+') changes only modestly with increasing $\beta$. This is consistent with the results of §3.2, which showed that moderate $\beta$ causes only a small geometric adjustment to the attached eddy hierarchy in the form of a change in the eddy aspect ratio ($AR$). The decomposition in figures 5(d–f) reaffirms this further; the wall-coherent contributions remain nominally unchanged with increasing $\beta$, and the peak of this high-energy region, particularly around the yellow '+', shows no appreciable shift. By contrast, the main effect of $\beta$ is the progressive emergence of an additional $\delta$-scaled contribution centred around $z/\delta \approx 0.3$, indicated by the magenta line. Not only are these motions smaller than the superstructures, but this APG-induced contribution is also predominantly wall-incoherent. We reaffirm this next by considering the integrated versions of the wall-coherent and incoherent spectra over $T^+$ following:

$$\overline{u^2}^+(z^+) = \underbrace{\int (\gamma_L^2)\, \phi_{uu}^+(z^+;\, T^+)\, \mathrm{d}T^+}_{\overline{u^2}^+_{\mathrm{WC}}:\ \text{wall-coherent}} + \underbrace{\int (1-\gamma_L^2)\, \phi_{uu}^+(z^+;\, T^+)\, \mathrm{d}T^+}_{\overline{u^2}^+_{\mathrm{WIC}}:\ \text{wall-incoherent}} . \tag{3.6}$$

Figures 5(g-i) compare $\overline{u^2}^+$ with $\overline{u^2}^+_{\mathrm{WIC}}$ and $\overline{u^2}^+_{\mathrm{WC}}$ for the ZPG, APG1 and APG3 cases, respectively. Also plotted for reference is the inverse log-law $\left(\overline{u^2}^+ = B_1 - A_1 \ln(z/\delta)\right)$ that is expected from contributions originating purely from the attached eddy hierarchy (Baars & Marusic 2020; Deshpande *et al.* 2021), or canonical wall-flows at extreme $\mathrm{Re}_\tau$ (*i.e.*, when the attached eddies statistically dominate the flow; Smits *et al.* 2011). These plots indicate that the increasing deviation of $\overline{u^2}^+$ from the inverse log-law behaviour, with increasing $\beta$, primarily arises from the enhanced contributions of $\overline{u^2}^+_{\mathrm{WIC}}$ that linearly superimpose onto $\overline{u^2}^+_{\mathrm{WC}}$. This interpretation is supported by the close alignment of the $\overline{u^2}^+_{\mathrm{WC}}$ profiles with the inverse log-law scaling, which exhibit a consistent slope of $A_1 = 0.98$ across all PG cases. Although this supports the approximate invariance of the attached-eddy contributions to $\overline{u^2}^+_{\mathrm{WC}}$, it is worth acknowledging the relatively minor yet non-negligible contributions from turbulent superstructures to this component. Per Baars & Marusic (2020) and Deshpande *et al.* (2021), a more rigorous estimate of $A_1$ requires removal of these superstructure contributions via a linear filtering methodology based on an outer-log reference location. Such a subsequent energy decomposition cannot be implemented here, however, owing to the limited physical extent of the logarithmic region.

While figure 5 shows the wall-coherent and incoherent energy distributions for each case separately, the baseline ZPG contributions related to the flow $\mathrm{Re}_\tau$, remain present in both components. To highlight the change introduced by APG, figures 6(a,b) present the premultiplied energy spectra subtracted between the APG and ZPG cases ($f\phi^+_{uu|\,\mathrm{APG-ZPG}}$), together with the corresponding subtracted wall-coherent and incoherent spectral subsets for the APG1 and APG3 cases. Negligible differences are noted at viscous and intermediate wall-coherent scales within the log region or below (*i.e.*, Regions I and II), with only small differences observed at scales associated with superstructures ($TU_\infty \approx 4.8\delta$) for the present low-to-moderate $\beta$ cases. Figures 6(c,d) show the same decomposition applied to the subtracted $\overline{u^2}^+$ profiles, with the reference markers retained from figure 5. Again, negligible differences are noted in $\overline{u^2}^+_{\mathrm{WIC}}$ in the log-region or below, which reinforces that the new APG-induced energy is predominantly a $\delta$-scaled contribution centred in the outer/wake region ($z/\delta \approx 0.3$). The fact that this energisation appears primarily in the wall-incoherent component is reaffirmed by the matching of the outer peaks of $\overline{u^2}^+|_{\mathrm{APG-ZPG}}$ and $\overline{u^2}^+_{\mathrm{WIC}}|_{\mathrm{APG-ZPG}}$. In contrast, the small yet non-negligible growth in $\overline{u^2}^+_{\mathrm{WC}}|_{\mathrm{APG-ZPG}}$ with increasing $\beta$ appears to be primarily associated with APG-induced energisation of the superstructure scales. This motivates future work aimed at explicitly extracting and quantifying turbulent superstructure contributions using established energy decomposition methodologies (Baars & Marusic 2020; Deshpande *et al.* 2021).

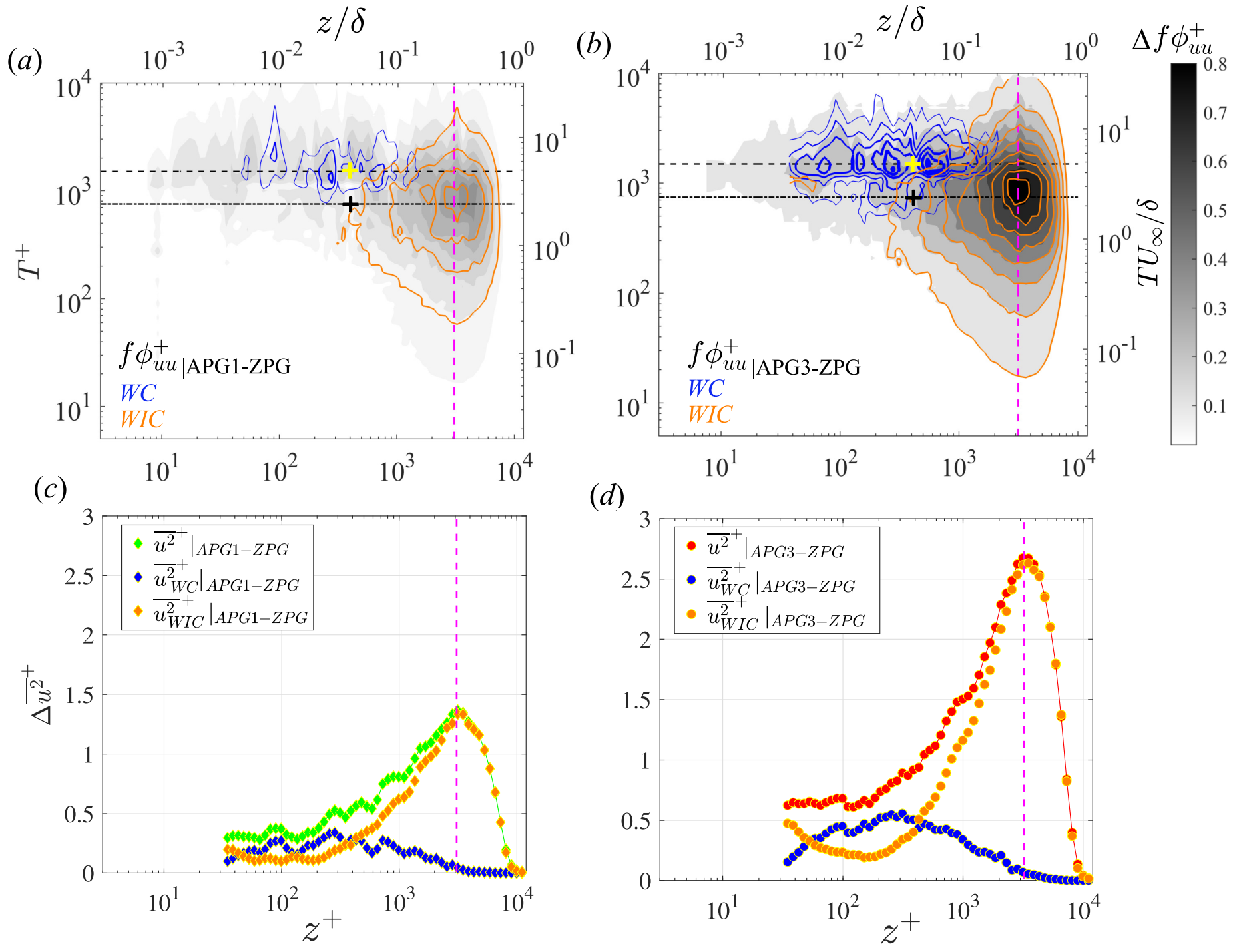

Figure 6: (a,b) Difference in premultiplied $u$-energy spectra corresponding to APG and ZPG cases, $f\phi^+_{uu|\,\mathrm{APG-ZPG}}$, considering the (a) APG1 and (b) APG3 cases. Also plotted are differences between spectra associated with the wall-coherent and incoherent contributions. (c,d) Corresponding differences between the $\overline{u^2}^+$ profiles and their wall-coherent and incoherent components for the (c) APG1 and (d) APG3 cases. Characteristic time scales associated with the APG-induced energy ($T^+/(U_\tau^2/U_\infty\nu) = 2.4\delta \approx 3\delta_{99}$) and superstructures ($T^+/(U_\tau^2/U_\infty\nu) = 4.8\delta \approx 6\delta_{99}$) are indicated in (a,b) using horizontal dashed and dash-dotted lines, while the vertical magenta line marks $z/\delta \approx 0.3$ in (a-d). The wall-coherent contribution is shown by blue contours plotted at identical levels for both APG1 and APG3 (0.1:0.04:0.3), with the line width increasing with contour level. The wall-incoherent contribution is plotted using contour levels of 0.1:0.1:0.8, which is the same for the grey-shading background used for the APG3 case.

In summary, the results of this section show that moderate APGs do not significantly modify the underlying hierarchy of wall-coherent motions established for canonical high-$\mathrm{Re}_\tau$ TBLs (regions I-III of figure 1). For instance, the near-wall cycle (region I) remains largely unchanged and the wall-coherent attached eddies/LSMs (region II) retain geometric self-similarity. The main effect of APG on this wall-coherent hierarchy is a modest geometric adjustment: the representative structures become relatively more inclined with respect to the wall and the wall-normal extent over which self-similarity is observed, is reduced. Additionally, VLSMs in the log region (region III) are only weakly energised, although this effect may become significant at higher $\beta$. The dominant APG effect is the introduction of $\delta$-scaled motions in the outer region (region IV). This contribution is primarily wall-incoherent and appears to linearly superimpose onto the pre-existing wall-coherent hierarchy of canonical ZPG TBLs.

## 4. Embedded shear layer dynamics in the TBL outer region

With the distinction between the APG-energised motions and the underlying wall-coherent hierarchy now established, we next examine the mechanisms/structures responsible for the APG-induced energy amplification within the wall-incoherent component in the outer region. As discussed in §1, several past studies (Bradshaw 1967; Elsberry *et al.* 2000; Kitsios *et al.* 2017; Schatzman & Thomas 2017; Silva & Wolf 2024) have noted a shear-layer-type flow instability mechanism that emerges in the outer region of strong APG TBLs. A necessary condition for any inviscid instability is the existence of an inflection point (IP) in the instantaneous velocity profile, $\widetilde{U}''(z_{\mathrm{IP}}) = 0$, where the prime denotes differentiation with respect to distance from the wall, $z$. Schatzman & Thomas (2017) noted two such inflection points in the streamwise *mean* velocity profiles of strong APG TBLs, indicating that the instantaneous streamwise velocity profiles are inflectional *on average*; namely, an inner (*i.e.*, close to the wall) and an outer inflection point. Consistent with this outer inflection point emerging in the mean velocity profile under strong APGs, it was hypothesised that their outer peak in the Reynolds shear stress profiles could be attributed to an increasingly energetic shear layer embedded in the TBL outer region. This hypothesis was further supported by the collapse of the mean turbulence statistics when shear-layer-inspired scalings were applied (Schatzman & Thomas 2017). This flow phenomenology is also supported by the experiments of Lozier *et al.* (2026*a*), where a synthetic shear layer structure was introduced into the outer region of a low-$\mathrm{Re}_\tau$ TBL; a brief discussion of their results has been included in appendix A for reference. In the case of low-to-moderate APG TBLs at high $\mathrm{Re}_\tau$, however, the outer peak in the Reynolds shear stress profile and the outer inflection point in the mean velocity profile are more subtle (figures 5i, 7a,c), requiring additional analyses and evidence to rigorously test this embedded shear layer hypothesis, as reported in this section.

Here, we evaluate the higher-order derivatives ($U'$ and $U''$) by employing the composite profile formulation for the mean velocity profile proposed in Zarei *et al.* (2026*b*), which provides a continuous profile and minimises the propagation of errors related to using discretely spaced data points along $z$. The Rayleigh–Fjørtoft criterion was then evaluated around each identified inflection point, $z_{\mathrm{IP}}$, per which the inviscid instability condition ($U''(U - U_{\mathrm{IP}}) < 0$) was satisfied only by the outer inflection point (*i.e.*, the instantaneous profiles are unstable around this point, on average). Figure 7(a) shows the detected outer inflection point for the APG3 profile around $z/\delta \approx 0.3$ which, albeit weak, is also consistent with the location of: (i) the maximum difference between the APG3 and ZPG spectra in figure 7(b) ($\Delta f \phi^{+}_{uu|\mathrm{APG3-ZPG}}$; same data as figure 6b), and (ii) the broad outer peak in the Reynolds shear stress (figure 7c). This motivates the use of conditional averaging to isolate dynamically significant events around the inflection point and the spectral peak, to assess if the moderately-strong APG-energised motions have shear-layer-type organisation/dynamics consistent with those noted in case of a strong APG. Inspired by Lee (2017), we condition the flow field based on the swirling-strength (described in §4.1) to identify intense spanwise vortical events within this region of interest, while §4.2 presents the conditional statistics used to test the shear-layer-type organisation of the APG-energised motions.

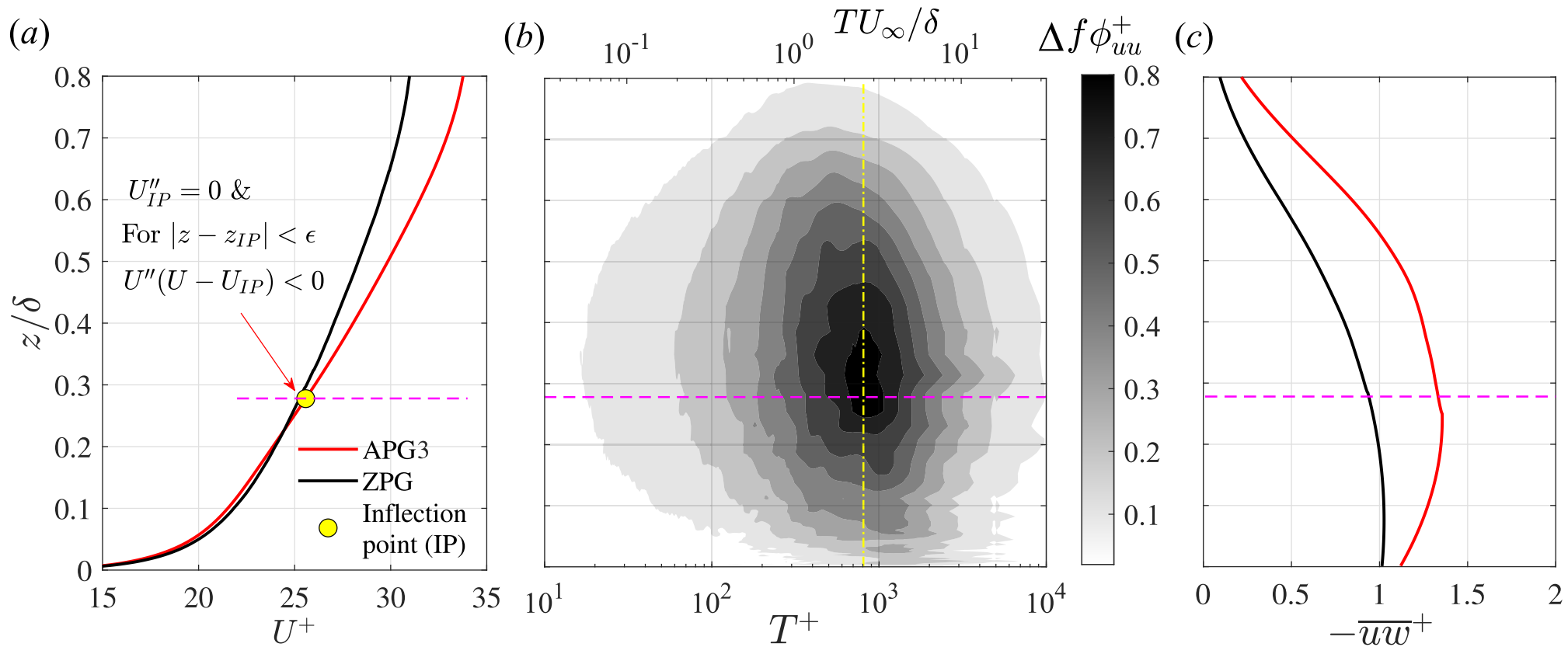

Figure 7: ($a$) Mean velocity profiles for the ZPG and APG3 cases. The yellow circle marks the detected outer inflection point for the APG3 case; no outer inflection point is observed for the ZPG case. ($b$) Subtracted premultiplied energy spectra, $\Delta f\phi^+_{uu} = f\phi^+_{uu|\mathrm{APG3-ZPG}}$. The dashed magenta line marks the wall-normal location of the peak in $\Delta f\phi^+_{uu}$, which is consistent with the outer inflection point location in ($a$), while the vertical dash-dotted line indicates the characteristic time scale, $TU_\infty \approx 2.4\delta$, of the APG-energised motions. ($c$) Reynolds shear stress profiles, $-\overline{uw}^+$, for the ZPG and APG3 cases.

### 4.1. *Conditional-averaging methodology*

Since these embedded shear layers are associated with coherent, spanwise-oriented vortices with predominantly clockwise rotation (Schatzman & Thomas 2017), the identification of such vortices/events in the $x$–$z$ plane forms the basis of the present conditional averaging methodology. To this end, vortices are identified using the swirling strength, $\lambda_{ci}$ (*i.e.*, the magnitude of the imaginary part of the complex-conjugate eigenvalues of the local velocity-gradient tensor), which quantifies the local intensity of vortex cores and has been widely adopted in the literature (Zhou *et al.* 1999; Wu & Christensen 2006; Christensen & Adrian 2001; Lee 2017). For the present planar PIV measurements in the $x$–$z$ plane, it is calculated from the two-dimensional velocity-gradient tensor as follows:

$$\lambda_{ci} = \begin{cases} \sqrt{-(a^2 + bc)}, & \text{if } a^2 + bc < 0, \\ 0, & \text{otherwise,} \end{cases} \tag{4.1}$$

where

$$a = \frac{\partial u}{\partial x}, \qquad b = \frac{\partial u}{\partial z}, \qquad c = \frac{\partial w}{\partial x}. \tag{4.2}$$

Since $\lambda_{ci}$ alone does not provide a sense of rotation, the swirl direction is assigned using the local sign of the spanwise vorticity (Wu & Christensen 2006; Dennis & Nickels 2011), resulting in a swirling-strength parameter:

$$\Lambda_{ci} = \lambda_{ci}(\omega_y/|\omega_y|). \tag{4.3}$$

For the present conditional averaging methodology, we retain only clockwise-rotating vortices by requiring a positive swirling-strength parameter, $\Lambda_{ci} > 0$ (given clockwise is defined as $\omega_y > 0$). The conditionally-averaged velocity fields are then calculated based on a linear stochastic estimation framework (Christensen & Adrian 2001; Lee 2017), yielding the conditional streamwise and wall-normal velocity components as functions of the relative

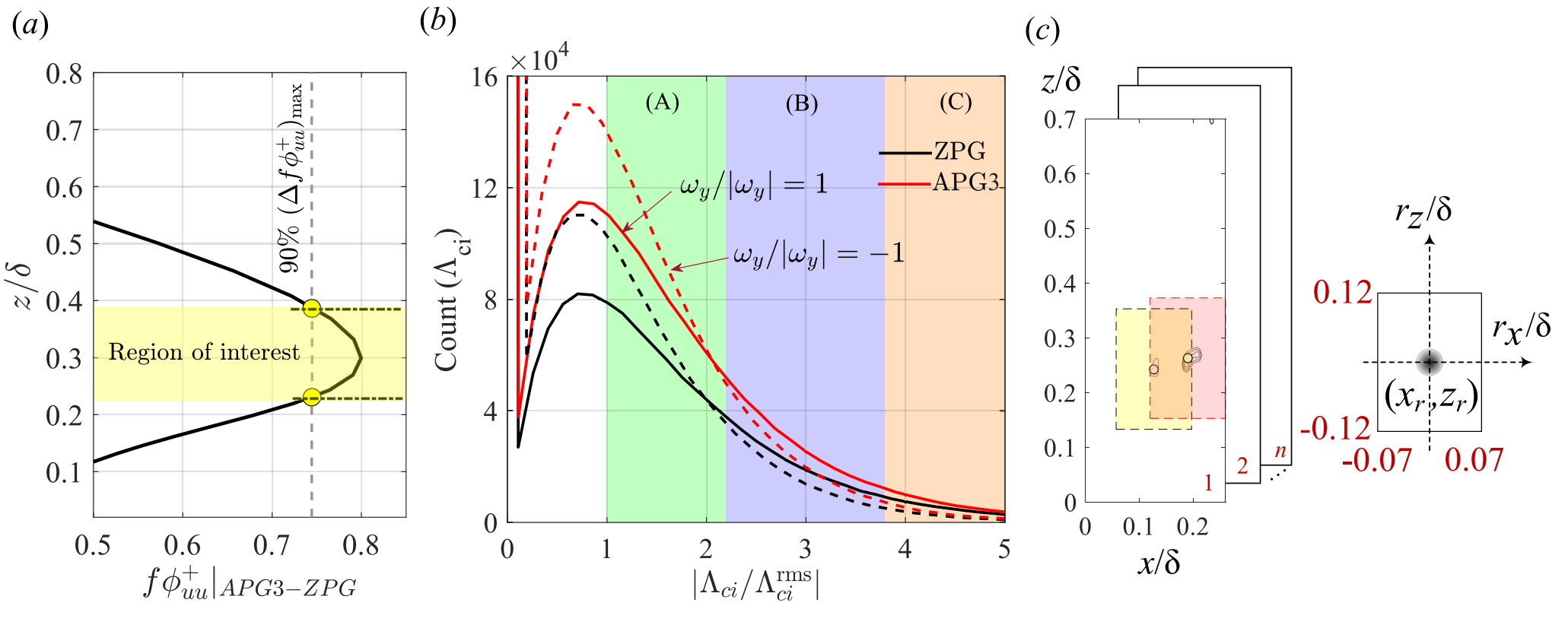

Figure 8: ($a$) Subtracted premultiplied energy spectra between APG3 and ZPG cases, $\Delta f\phi^+_{uu}$, at the time scale of the APG-energised peak, $TU_\infty \approx 2.4\delta$ (vertical dashed line in figure 7b). Shaded region represents $z/\delta \approx 0.2$–$0.4$ where $\Delta f\phi^+_{uu}$ is above 90% of the maximum. ($b$) Probability density functions of the normalised swirling-strength parameter within the region of interest, $|\Lambda_{ci}/\Lambda^{\mathrm{rms}}_{ci}|$; shaded regions A–C are used to guide the threshold selection for dynamically significant events. ($c$) Schematic of the conditional averaging procedure showing example reference points and the extracted boxes used for ensemble averaging.

spatial coordinates $(r_x, r_z)$:

$$\langle u_j\rangle(r_x, r_z) = \frac{\{\Lambda_{ci}(x_r, z_r)\, u_j(x_r + r_x,\, z_r + r_z)\}}{\{\Lambda^2_{ci}(x_r, z_r)\}}\, \Lambda_{ci}(x_r, z_r). \qquad u_j \in (u, w) \qquad (4.4)$$

Here, $(x_r, z_r)$ denotes the location of the identified vortex core and serves as the reference point for the conditional averaging. As shown in figure 8(a), we first define the wall-normal region of interest for conducting this conditional averaging by considering the difference between the premultiplied spectra of the APG3 and ZPG TBLs at the time scale corresponding to the APG-energised peak ($TU_\infty \approx 2.4\delta$). Considering the region above 90% of the maximum to be of most interest, we identify $0.2 < z/\delta < 0.4$ to correspond with the dominant APG-energised structures and assign this range for the conditional averaging (referred henceforth as the 'wall-normal region of interest'). Although the signature of shear-layer-type motions may extend beyond this range, we expect the associated vortices to be strongest within this wall-normal range per previous observations (Schatzman & Thomas 2017; Deshpande & Vinuesa 2024).

In practice, a threshold ($\Lambda_{ci,th}$) is also required to distinguish significant vortices/events from weaker background fluctuations. However, the precise choice of threshold is subjective and depends on multiple factors, including the spatial resolution of the dataset, the measurement noise level, and the objective of the analysis. For this reason, different threshold definitions can be found in the literature, based on either the maximum ($\Lambda^{\max}_{ci}$) or the root-mean-square ($\Lambda^{\mathrm{rms}}_{ci}$) of the swirling-strength parameter. For example, Dennis & Nickels (2011) used $\Lambda_{ci,th} = 0.18\,\Lambda^{\max}_{ci}$, whereas Wu & Christensen (2006) and Parthasarathy & Saxton-Fox (2025) used $\Lambda_{ci,th} = 1.5\,\Lambda^{\mathrm{rms}}_{ci}$. In the present study, we establish an empirical threshold based on the relative populations/contributions of clockwise- and counter-clockwise-oriented vortices. To this end, figure 8(b) shows the probability density functions of $\Lambda_{ci}$ extracted from the wall-normal region of interest for both the ZPG and APG3 cases, plotted as a function of $|\Lambda_{ci}/\Lambda^{\mathrm{rms}}_{ci}|$. Consistent with the observation of Lee *et al.* (2010), higher probability density values indicate larger populations and/or stronger vortices in the APG3 case relative to the ZPG case, across the full range of $\Lambda_{ci}$. Interestingly, while the clockwise-rotating vortices dominate at higher-magnitudes of $\Lambda_{ci}$, counter-clockwise swirl dominates for $|\Lambda_{ci}/\Lambda_{ci}^{\mathrm{rms}}| \lesssim 2.2$. These critical bandwidths are reflected through the shaded regions, A, B and C, indicated by background colours in figure 8(b). While each of these ranges was tested, we selected $\Lambda_{ci,th} \sim 3.8\ \Lambda_{ci}^{\mathrm{rms}}$ (corresponding to region C) as the threshold for identifying statistically significant clockwise vortices in the present study. Detailed sensitivity analyses for the thresholds defined based on regions A−C, their corresponding conditional averaging results, and further justification for the chosen threshold has been documented in Lozier *et al.* (2026*b*).

This threshold was then applied within the wall-normal region of interest for each PIV flow field snapshot to detect significant vortices/events. A sufficiently large box ($\Delta x \times \Delta z \sim 0.14\delta \times 0.24\delta$) was also defined in order to span the region of interest and capture both the vortex core and its surrounding velocity field. Figure 8(c) shows a schematic of two such vortices/events detected in sample PIV fields 1 and 2, and the corresponding boxes centred at $(x_r, z_r)$ together with the box dimensions. The conditional fields were then obtained by ensemble averaging over all such accepted events, and across all snapshots, with each event contributing one box-sized velocity field centred at $(x_r, z_r)$. The same threshold definition was used for both the ZPG and APG3 cases to enable direct comparison of the statistics.

### 4.2. *Conditionally-averaged flow structure and dynamics*

We now utilise the above methodology to compute conditionally-averaged flow fields, as shown in figure 9, and examine if they exhibit characteristics consistent with an embedded shear layer. Figures 9(*a–c*), respectively, depict conditional velocity vector fields overlaid with $\langle\Lambda_{ci}\rangle^+$, conditional streamwise velocity fluctuations ($\langle u\rangle^+$) and conditional wall-normal velocity fluctuations ($\langle w\rangle^+$) for the ZPG case, with the corresponding fields for the APG3 case shown in figures 9(*d–f*). Compared with the ZPG case, the APG3 case exhibits substantially stronger $\langle\Lambda_{ci}\rangle^+$ as well as a much clearer clockwise rotation in the conditional velocity vectors around the vortex core reference point. This is accompanied by relatively larger amplitudes and spatial extents of $\langle u\rangle^+$ and $\langle w\rangle^+$ compared to the ZPG case. The conditional streamwise fluctuations show that $\langle u\rangle^+ > 0$ for $r_z/\delta > 0$ and $\langle u\rangle^+ < 0$ for $r_z/\delta < 0$, indicating that the conditional event is associated with a high- and low-momentum region above and below the reference point, respectively. Meanwhile, the conditional wall-normal fluctuations show that $\langle w\rangle^+ > 0$ for $r_x/\delta < 0$ and $\langle w\rangle^+ < 0$ for $r_x/\delta > 0$, consistent with shear-layer-type motions around the reference point (Schatzman & Thomas 2017). This spatial distribution of $\langle u\rangle^+$ and $\langle w\rangle^+$, and their increasing amplitudes under APG, are also consistent with the conditional velocity signatures reported in previous APG studies (Volino *et al.* 2007; Lee & Sung 2009; Lee 2017). The same general patterns of conditional motions in the ZPG case, but with markedly weaker amplitudes, is likely associated with the statistically weaker shear layers between uniform momentum zones of canonical TBLs (Bautista *et al.* 2019; Gul *et al.* 2020). Whether the stronger shear-layer-type signatures noted for the APG case are simply energised equivalents of those existing in a ZPG TBL, or are associated with a new coherent flow structure, would likely require a detailed three-dimensional flow characterisation that remains beyond the scope of the present study.

Figures 10(a,b) then show the conditionally-averaged Q2 and Q4 contributions to the Reynolds shear stress, $-\langle uw\rangle^+$, for the ZPG and APG3 cases, while figure 10(c) shows the corresponding conditionally-averaged mean streamwise velocity profiles ($\langle U\rangle^+ - \langle U\rangle^+_{\mathrm{avg}}$) to establish a connection between the mean velocity profile and the shear-layer-type motions in the outer region. Here, the profile of $\langle U\rangle^+$ is extracted at $r_x = 0$ and the subscript 'avg' denotes averaging across the full range of $r_z$. In the APG3 case, a clear dipolar organisation of Q2 and Q4 events is observed, where Q2 corresponds to $\langle u\rangle < 0$, $\langle w\rangle > 0$ (ejections) and Q4 to $\langle u\rangle > 0$, $\langle w\rangle < 0$ (sweeps). This arrangement is consistent with spanwise-oriented clockwise rollers, as noted by Schatzman & Thomas (2017), confirming that embedded-shear-layer-type

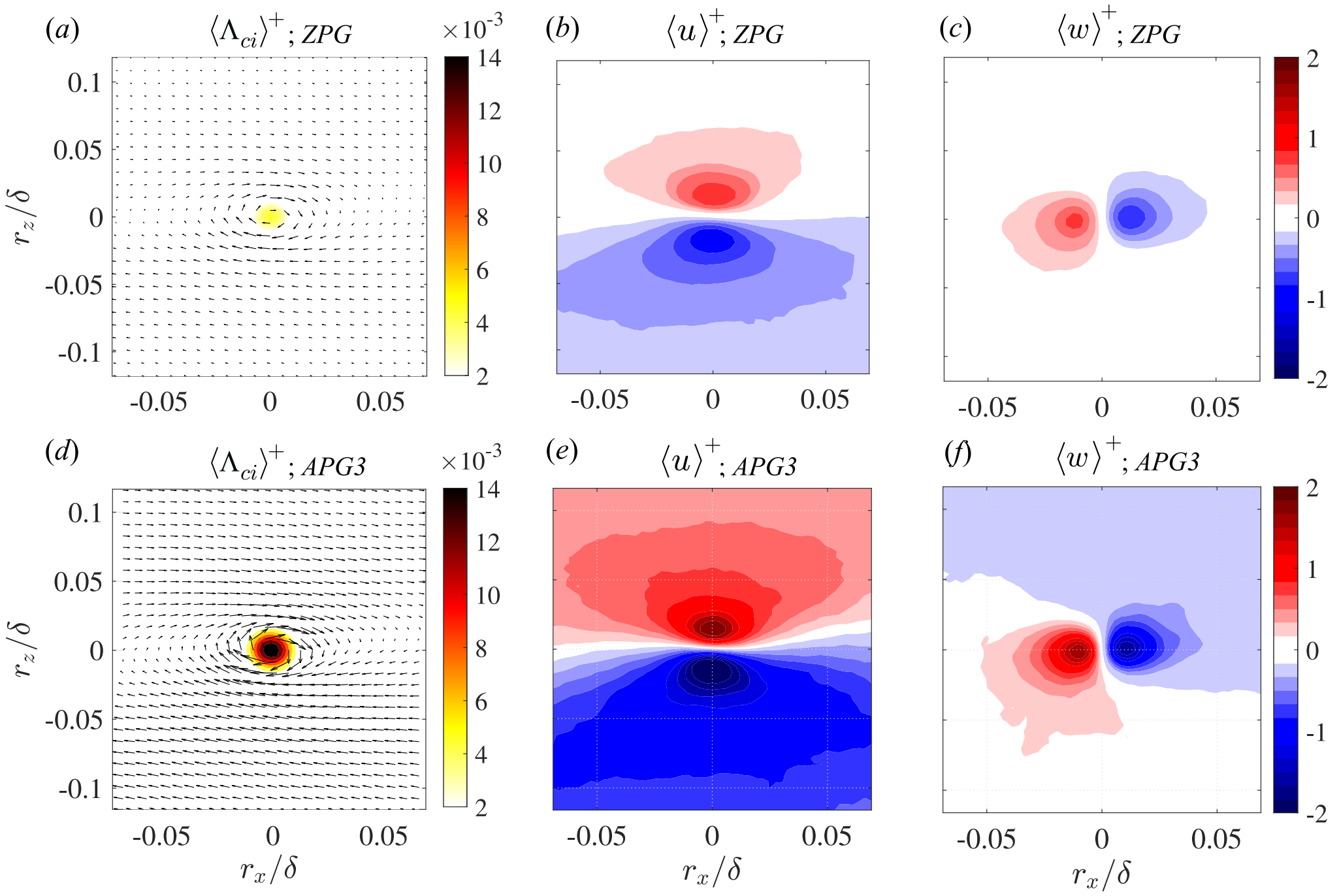

Figure 9: Conditionally-averaged flow fields conditioned on $\Lambda_{ci} \gtrsim 3.8\ \Lambda_{ci}^{\mathrm{rms}}$ for $0.2 \lesssim z/\delta \lesssim 0.4$. (a,d) Conditionally-averaged velocity vectors and swirling strength $\langle\Lambda_{ci}\rangle^+$, (b,c,e,f) conditionally-averaged streamwise and wall-normal velocity fluctuations. (a-c) correspond to the ZPG case, while (d-f) correspond to the APG3 case. The relative coordinates $(r_x, r_z)$ are defined with respect to vortex cores at $(r_x, r_z) = (0, 0)$.

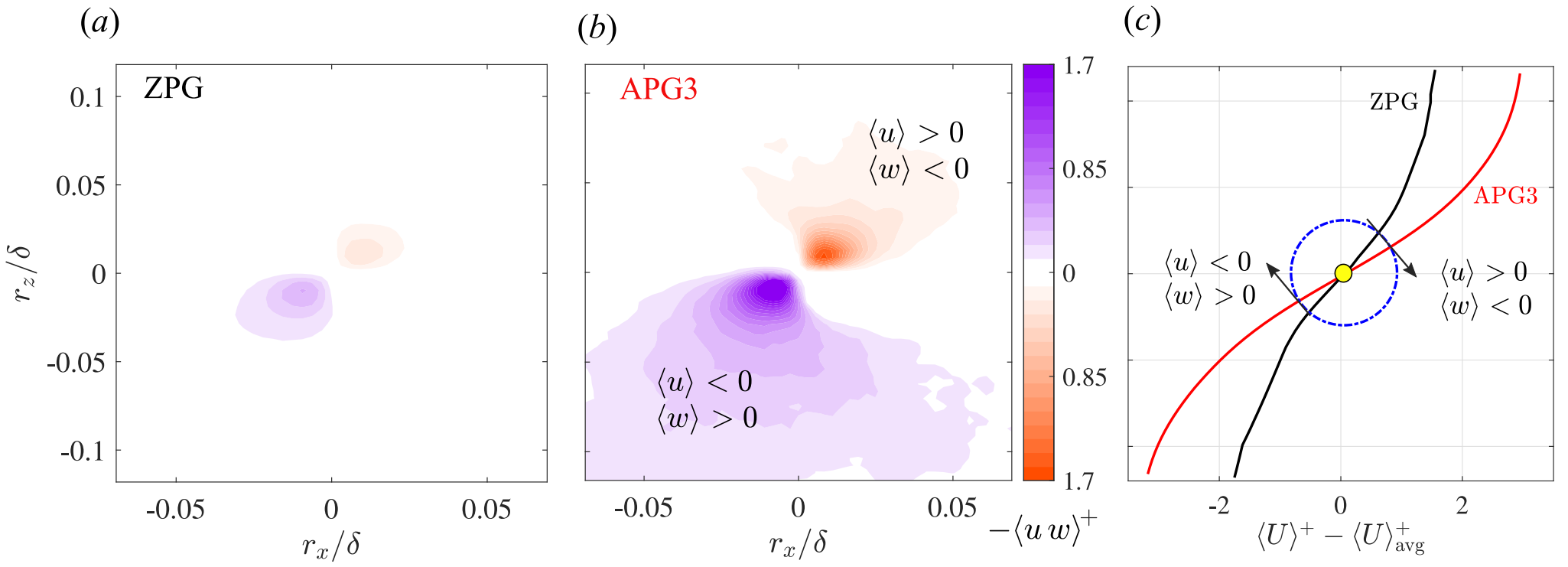

Figure 10: Conditionally-averaged Q2 and Q4 contributions to $-\langle uw\rangle^+$ for the (*a*) ZPG and (*b*) APG3 cases. (*c*) Conditionally-averaged mean streamwise velocity profiles, $\langle U\rangle^+$ subtracted from the $\langle U\rangle^+_{\mathrm{avg}}$.

motions are not limited to strong APG TBLs. The elevated values of both $-\langle uw\rangle^+$ and $\langle u^2\rangle^+$ in the outer region, relative to ZPG case, are also consistent with the conceptual picture presented by Deshpande & Vinuesa (2024). The conditionally-averaged velocity profile of the APG3 cases also exhibits a pronounced (and recognisably unstable) inflectional shape, compared to the ZPG case, directly connecting the contributions of these specific conditional vortices to the inflectional outer mean velocity profile plotted in figure 7(a).

Additionally, a representative Strouhal number, $St$, can be estimated from the APG3 velocity profile in figure 10(c), where $St = f_s\theta_d/\langle U\rangle_{\mathrm{avg}}$ (Ho & Huerre 1984). Here, $f_s$ is the representative frequency of the APG-energised motions and $\theta_d$ is the momentum thickness evaluated from the conditionally-averaged velocity profile $\langle U\rangle$. Rather than selecting a single value of $f_s$, we consider the frequency (time-scale) range from figure 7(b) over which the difference between the APG and ZPG spectra remains above 80% of its maximum value ($1.5 < TU_\infty/\delta < 6.0$). Based on this representative frequency range, we estimate that the Strouhal number of these conditional, APG-energised vortices, within our region of interest, falls between $0.006 < St < 0.025$. Remarkably, this range corresponds to $St$ typically reported for shear layers (Schatzman & Thomas 2017), reaffirming the dynamic similarity between these APG-energised motions and shear-layer-type motions.

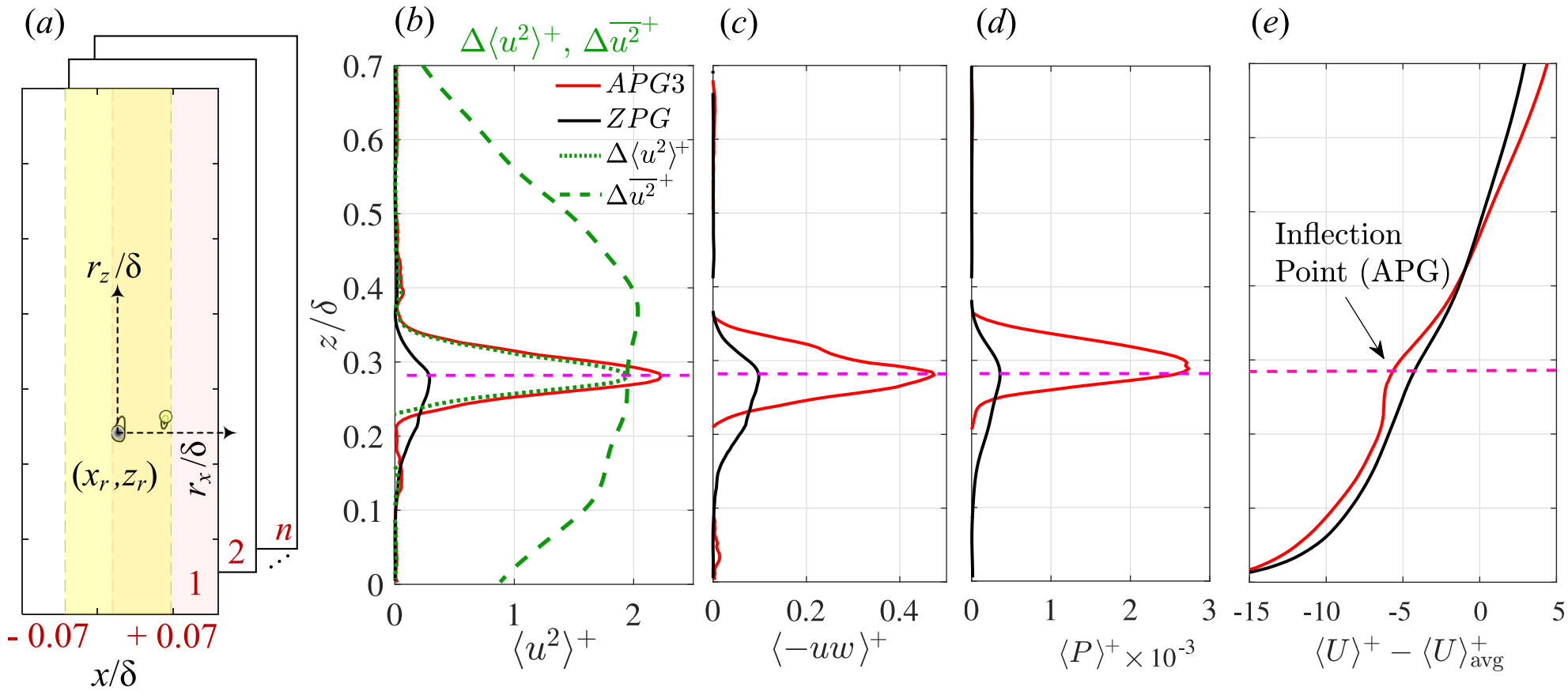

Figure 11: ($a$) Schematic of the event-centred conditioning boxes spanning the full boundary-layer thickness. Conditionally-averaged wall-normal profiles of the ($b$) streamwise TKE, ($c$) Reynolds shear stress, ($d$) TKE production and ($e$) mean streamwise velocity for the ZPG and APG3 cases. In (b), the green dashed and dotted lines, respectively, show the differences between APG3 and ZPG cases for the conventional $\left(\Delta\overline{u^2}^+ = \overline{u^2}^+_{\mathrm{APG3}} - \overline{u^2}^+_{\mathrm{ZPG}}\right)$ and conditionally-averaged $\left(\Delta\langle u^2\rangle^+ = \langle u^2\rangle^+_{\mathrm{APG3}} - \langle u^2\rangle^+_{\mathrm{ZPG}}\right)$ streamwise TKE profiles. The magenta dashed line indicates the wall-normal location for the outer inflection point in $\langle U\rangle^+$.

Next, conditional averaging is performed using the same $\Lambda_{ci}$ threshold, but with an extended sampling box, spanning the full boundary-layer thickness, such that the dynamically significant clockwise-rotating spanwise vortices are retained at their original wall-normal locations within $0.2 < z/\delta < 0.4$ (see figure 11a). This enables assessment of their combined contributions to the wall-normal profiles of conventional one-point turbulence statistics ($\langle u^2\rangle^+$, $\langle -uw\rangle^+$, $\langle U\rangle$ and the TKE production $\langle P\rangle^+$) as plotted in figures 11(b–e). Here, the conditionally-averaged TKE production is evaluated based on both the conditionally-averaged mean shear and Reynolds shear stress, *i.e.*, $\langle P\rangle^+ \equiv -\langle uw\rangle^+\,\partial\langle U\rangle^+/\partial z^+$. The increase in all conditionally-averaged statistics for the APG3 case, aligned with the outer inflection point (figure 11e), supports the interpretation that these APG-energised outer-region motions have shear-layer-type organisation/dynamics, consistent with the hypothesis of Schatzman & Thomas (2017). Figure 11(b) also shows the difference between the APG3 and ZPG cases for the conditionally-averaged $\langle u^2\rangle^+$, compared against the difference between $\overline{u^2}^+$. The agreement between these quantities around $z \sim 0.3\delta$ confirms that a majority of the additional streamwise TKE introduced by the APG ($\Delta\overline{u^2}^+$) is accounted for, locally, by the embedded-shear-layer-type motions ($\Delta\langle u^{2+}\rangle$). While the present conditioning only considers intense vortices/events within $0.2 < z/\delta < 0.4$, application of the same procedure across neighbouring wall-normal regions is expected to lead to similar conclusions.

In summary, the results of this section support the hypothesis that the APG-induced energy amplification in region IV of figure 1(a) is associated with the energisation of embedded-shear-layer-type motions in the outer region of the TBL. These motions are localised near the outer inflection point and exhibit characteristic signatures of shear-layer dynamics, including intensified spanwise vortices, stronger sweep and ejection events, and increased Reynolds shear stress and TKE production relative to ZPG TBLs. The observed conditional velocity organisation and representative Strouhal numbers are also consistent with previous observations of typical shear layers. Together, these results indicate that embedded shear layers provide a plausible physical mechanism underlying the dominant wall-incoherent energy introduced by moderate APG, through the energisation of an outer-region dynamical pathway that also exists, albeit weakly, in canonical ZPG TBLs.

## 5. Summary and concluding remarks

The present study experimentally characterises APG-energised motions/eddies in the outer region of high-$\mathrm{Re}_\tau$ TBLs. All experiments were conducted under minimal upstream PG history conditions, and at sufficiently high $\mathrm{Re}_\tau$, to enable an unambiguous assessment of the local APG effect and ensure sufficient separation between inner- and outer-scaled motions, respectively. A new two-point hot-wire dataset was used to analyse and compare the time-scale-dependent linear coherence of streamwise TKE-carrying motions between a ZPG and moderate-APG TBL at matched $\mathrm{Re}_\tau$. Conditional averaging was also performed on a matched, previously published, PIV dataset (Lindić *et al.* 2025) to quantify the distribution of these APG-energised eddies and their contributions to the mean turbulence statistics.

Figures 12(a,b) present a conceptual picture summarising the main findings on the scaling and organisation of streamwise TKE-carrying motions, while also inviting discussion about their implications on the classical logarithmic scalings of $\overline{u^2}^+$ and $U^+$ summarised in figures 12(c,d) (Baars & Marusic 2020; Zarei *et al.* 2026*a*). The LCS analysis revealed that the hierarchy of wall-coherent motions, which comprises the viscous-scaled near-wall cycle, $z$-scaled attached eddy hierarchy and $\delta$-scaled superstructures, all exhibit consistent scaling and energy distribution across the ZPG and APG cases (figures 5,6). The only notable differences were: (i) the slight energisation of the superstructures and (ii) the gradual reduction in the streamwise–wall-normal aspect ratio of the $z$-scaled attached eddies with increasing $\beta$; the latter was related to the increase in the mean inclination angle and mean wall-normal velocity (figure 4). The most significant difference between the ZPG and moderate APG TBL was associated with the subset of the streamwise TKE that is *incoherent* with the wall ($\overline{u^2}^+_{\mathrm{WIC}}$; figure 6). While there were negligible differences in $\overline{u^2}^+_{\mathrm{WIC}}$ within the log region, the increase in the wake/outer region accounted for almost all the streamwise TKE added on imposition of the APG. As such, this scale-specific energy decomposition methodology confirmed that the majority of the APG-induced streamwise TKE is linearly superimposed onto the wall-coherent hierarchy of motions ($\overline{u^2}^+_{\mathrm{WC}}$). This superimposed energy is then responsible for the following observations related to increasing $\beta$: (i) increasing deviation of the $\overline{u^2}^+$ profile from the inverse log-law scaling ($\overline{u^2}^+ = B_1 - A_1\ln(z/\delta)$) associated with the wall-coherent attached eddy hierarchy (figure 5), and (ii) lowering the upper bound of the $U^+$ log-law scaling thereby diminishing the wall-normal range exhibiting self-similarity. Nevertheless, the exhibition of consistent scalings by $\overline{u^2}^+_{\mathrm{WIC}}$ and $\overline{u^2}^+_{\mathrm{WC}}$, between the ZPG and APG cases, strongly suggests applicability of the data-driven modelling framework of Baars & Marusic (2020) for predicting $\overline{u^2}^+$ as a function of $\mathrm{Re}_\tau$ and $\beta$, which we leave for future works.

While the wall-coherent streamwise TKE has been associated in the literature with distinct coherent structure types (figure 1), no such association has been established for the wall-incoherent energy in the outer region. Here, we tested the applicability of the embedded shear layer hypothesis of Schatzman & Thomas (2017) for describing the structure of moderately-strong APG-energisation in the outer region (which is predominantly wall-incoherent). This was achieved by conditionally-averaging the 2-D $x$-$z$ PIV flow fields based on intense spanwise swirling events (*i.e.*, clockwise vortices). The conditional statistics established a direct connection between APG-induced energisation centred at $z/\delta \approx 0.3$, with the outer inflectional point of the mean velocity profile, outer peak in the TKE production and a clear Q2-Q4 organisation of Reynolds shear stress events consistent with shear-layer-type motions (Schatzman & Thomas 2017; Deshpande & Vinuesa 2024). Furthermore, the characteristic Strouhal number obtained from the conditionally-averaged APG TBL velocity profile ($\langle U \rangle$) was found to lie within the range reported previously for shear layers (Schatzman & Thomas 2017), supporting the presence of shear layer dynamics in the outer region of APG TBLs. Taken together, these findings suggest that the principal effect of APG is not to fundamentally alter the attached-eddy hierarchy, but rather to modulate the energetic competition between two coexisting dynamical pathways in the outer-region: a wall-coherent pathway associated with the attached eddy hierarchy and superstructures, and a wall-incoherent pathway associated with shear-layer-type motions. Increasing APG progressively strengthens the latter pathway, allowing outer-scaled motions to contribute more significantly to the net turbulence statistics.

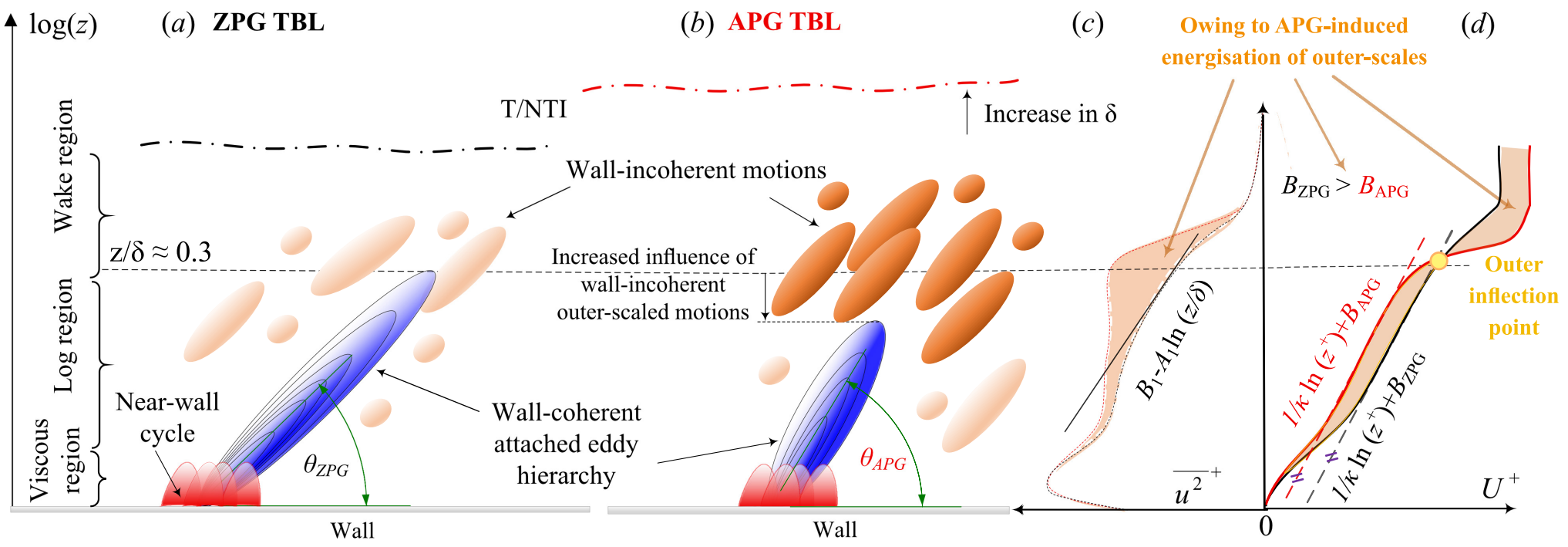

Figure 12: Conceptual sketch highlighting differences in the streamwise TKE-carrying energetic motions between a high-$Re_\tau$ (a) ZPG and (b) low-to-moderately strong APG TBL. While the viscous-scaled near-wall (in red) and the wall-scaled attached eddy hierarchy (in blue) exhibits similar scaling in both ZPG and low-to-moderate APG TBLs, the attached eddies are found to incline at a steeper angle (*i.e.*, $\theta_{APG} > \theta_{ZPG}$). The most significant differences are noted for the wall-incoherent motions in the outer/wake region that exhibit outer-scaling, which get significantly energised and penetrate from the top of the log region. (c,d) This explains the deviations from the log-law scalings of $\overline{u^2}^+$ associated with the attached eddies (blue trapezoid in (c)) and $U^+$ (dashed lines in (d)) noted for a ZPG TBL (Baars & Marusic 2020; Zarei *et al.* 2026*a*).

The present study, however, is limited in its scope, given that it cannot distinguish whether these shear-layer-type motions represent new structures introduced in the APG TBL by the flow instability (related to the outer inflection point) or simply an amplified manifestation of the shear layers coexisting between uniform momentum zones (also present in a ZPG TBL, albeit weakly). We hypothesise that the latter scenario is more likely, given that the mean streamwise velocity profiles of both ZPG and APG TBLs have been successfully described through a single composite profile formulation (Coles 1956; Nickels 2004; Zarei *et al.* 2026*a*) that models the outer region as a wake (*i.e.*, conforming to shear layer behaviours). This hypothesis also aligns with the observations of Gungor *et al.* (2022) and Deshpande &

Vinuesa (2024), who noted similarity in the fundamental energy transfer mechanisms in the outer regions of ZPG and APG TBL flows. The same hypothesis could also be extrapolated to explain observations in the literature related to the favourable-pressure-gradient (FPG) TBLs (Jones *et al.* 2001). If imposition of an APG strengthens the outer-scaled wall-incoherent contributions that penetrate deeper into the log region, then imposition of FPG condition may have the opposite effect; *i.e.*, the wall-incoherent outer-scaled contribution may weaken more than in the case of ZPG, leading to the extension of the wall-coherent attached eddy hierarchy over a larger fraction of the TBL. In the limiting equilibrium case of the sink flow, this could correspond to the attached eddy hierarchy extending towards the boundary layer edge, explaining the exhibition of the classical logarithmic scalings for $\overline{u^2}^+$ and $U^+$ across an extended fraction of the TBL (relative to ZPG case; Jones *et al.* 2001). These hypotheses motivate further systematic investigation of high-$Re_\tau$ PG TBLs in the future for understanding the interplay between the attached eddy hierarchy and outer-scaled wake structures for varying PG strengths/conditions.

**Acknowledgements.** I. Marusic and R. Deshpande gratefully acknowledge funding from the Office of Naval Research (ONR) and ONR Global; Grant No. N62909-23-1-2068. R. Deshpande is also grateful for financial support from the Melbourne Postdoctoral Fellowship and the Vice Chancellor's Fellowship from the University of Melbourne and RMIT University, respectively.

**Declaration of Interests.** The authors report no conflict of interest.

## Appendix A. Artificially introduced shear layer in the TBL outer region

To further support the interpretation that APG-induced changes in the mean turbulence statistics are linked to the energisation of embedded shear-layer-type structures, we consider a series of experiments in which *non-native* shear-layer motions were synthetically introduced into the outer region of a low-$Re_\tau$ ZPG TBL through active flow control. These experiments were performed in a low-turbulence, subsonic, in-draft wind tunnel at the Hessert Laboratory for Aerospace Research at the University of Notre Dame (Lozier *et al.* 2026*a*) where a plasma-based actuator was mounted on the top surface of a ZPG boundary-layer development plate near the downstream end of the test section (figure 13*a*). Alternating-current dielectric barrier discharge (AC-DBD) plasma was generated by connecting the copper electrodes on the top and bottom of the dielectric actuator plate to a high-voltage AC source, producing a quasi-steady plasma jet above the actuator plate's top surface. The jet was then sinusoidally modulated to impose a strong periodic perturbation to the local mean shear, thereby introducing large-scale, spanwise-uniform motions into the outer region of the TBL and generating an inflectional mean velocity profile downstream of the actuator, thereby reproducing the key kinematic condition associated with shear-layer instability in APG TBLs.

Figure 13(b) shows the mean velocity at $x = 1.5\delta_0$ (with $\delta_0$ denoting the canonical boundary-layer thickness at $x = 0$), where the forced profile develops an outer inflection point at $z = H$, with $H$ corresponding to the wall-normal location of the plasma actuator. Phase-locked PIV measurements were acquired of the two-dimensional velocity field over a long streamwise extent downstream of the actuator. Conditional flow statistics were then obtained using the same $\Lambda_{ci}$-based averaging criterion and threshold adopted in the present study. Figure 13(c) shows the conditional Q2-Q4 contributions to $-\langle uw\rangle^+$ together with the $\Lambda_{ci}$ contour, demonstrating the same organised Q2–Q4 signature associated with spanwise-oriented shear-layer rollers observed in the APG TBL, while figure 13(d) shows the corresponding conditional velocity field and spanwise vorticity, confirming the coherent organisation of the imposed motions. Interestingly, further downstream, figure 13(e) shows that at $x = 5\delta_0$ the normalised mean velocity exhibits a negative shift in the log region and a strengthened wake relative to the baseline ZPG case, consistent with trends reported for APG TBLs (Nagano *et al.* 1998; Deshpande *et al.* 2023*b*). Overall, these results reproduce the key conditional signatures observed in §4, namely the inflectional mean profile, organised Q2–Q4 structure, and outer-region energisation. This provides further evidence that embedded shear-layer-type motions are sufficient to produce turbulence-statistical signatures closely resembling those observed in APG TBLs.

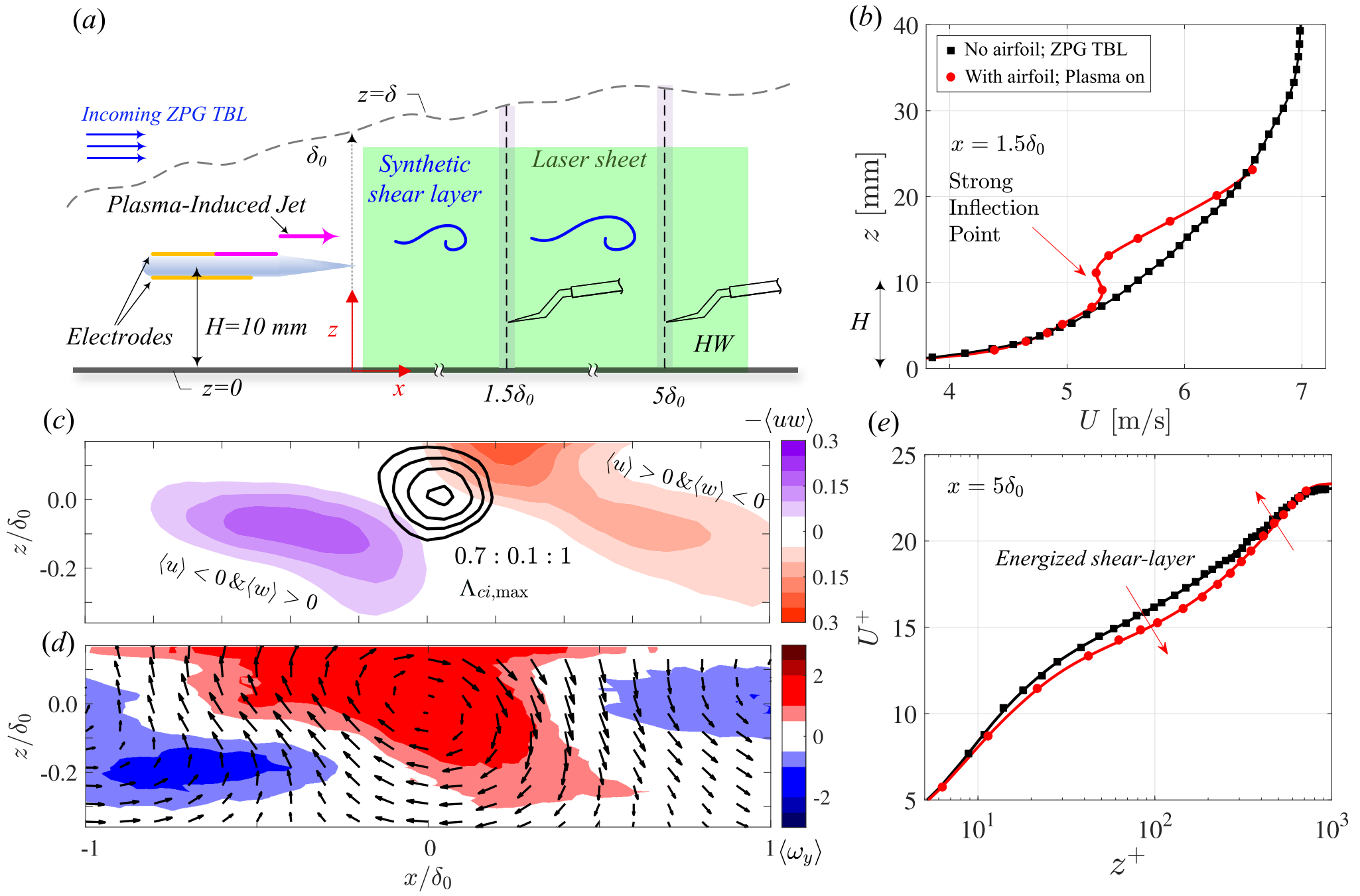

Figure 13: ($a$) Schematic of the experimental setup and plasma actuator arrangement used to introduce (non-native) shear-layer-type motions into the outer region of a low-$Re_\tau$ ZPG TBL (Lozier *et al.* 2026*a*). ($b$) Mean velocity profiles at $x = 1.5\delta_0$ for the baseline ZPG case and the plasma-on case, where $\delta_0$ is the canonical TBL thickness at $x = 0$. ($c$) Conditional Q2-Q4 contributions to $-\langle uw\rangle^+$ with $\Lambda_{ci}$ contours overlaid (shown for $\Lambda_{ci}/\Lambda_{ci,\max} = 0.7 : 0.1 : 1$). ($d$) Conditional velocity field and spanwise vorticity, highlighting the coherent organisation of the induced motions. ($e$) Mean velocity profiles at $x = 5\delta_0$ for ZPG and the plasma-on case, showing a downward shift in the plasma-on (forced) mean velocity profile.

## REFERENCES

Adrian, R. J., Meinhart, C. D. & Tomkins, C. D. 2000 Vortex organization in the outer region of the turbulent boundary layer. *J. Fluid Mech.* **422**, 1–54.

Aubertine, C. D. & Eaton, J. K. 2005 Turbulence development in a non-equilibrium turbulent boundary layer with mild adverse pressure gradient. *J. Fluid Mech.* **532**, 345–364.

Baars, W. J. & Marusic, I. 2020 Data-driven decomposition of the streamwise turbulence kinetic energy in boundary layers. Part 1. Energy spectra. *J. Fluid Mech.* **882**, A25.

Balakumar, B. J. & Adrian, R. J. 2007 Large- and very-large-scale motions in channel and boundary-layer flows. *Phil. Trans. R. Soc. A: Math. Phys. Engng Sci.* **365** (1852), 665–681.

Bautista, J. C. C ., Ebadi, A., White, C. M., Chini, G. P. & Klewicki, J. C. 2019 A uniform momentum zone–vortical fissure model of the turbulent boundary layer. *J. Fluid Mech.* **858**, 609–633.

Bendat, J. S. & Piersol, A. G. 2011 *Random Data: Analysis and Measurement Procedures*. Wiley.

Bobke, A., Vinuesa, R., Örlü, R. & Schlatter, P. 2017 History effects and near equilibrium in adverse-pressure-gradient turbulent boundary layers. *J. Fluid Mech.* **820**, 667–692.

Bradshaw, P. 1967 The turbulence structure of equilibrium boundary layers. *J. Fluid Mech.* **29**, 625–645.

Bross, M., Fuchs, T. & Kähler, C. J. 2019 Interaction of coherent flow structures in adverse pressure gradient turbulent boundary layers. *J. Fluid Mech.* **873**, 287–321.

Brown, G. L. & Thomas, A. S. W. 1977 Large structure in a turbulent boundary layer. *Phys. Fluids* **20** (10), S243–S252.

Christensen, K. T. & Adrian, R. J. 2001 Statistical evidence of hairpin vortex packets in wall turbulence. *J. Fluid Mech.* **431**, 433–443.

Clauser, F. H. 1954 Turbulent boundary layers in adverse pressure gradients. *J. Aeronaut. Sci.* **21** (2), 91–108.

Coles, D. 1956 The law of the wake in the turbulent boundary layer. *J. Fluid Mech.* **1** (2), 191–226.

Dennis, D. J. C. & Nickels, T. B. 2011 Experimental measurement of large-scale three-dimensional structures in a turbulent boundary layer. part 1. vortex packets. *J. Fluid Mech.* **673**, 180–217.

Deshpande, R., Monty, J. P. & Marusic, I. 2021 Active and inactive components of the streamwise velocity in wall-bounded turbulence. *J. Fluid Mech.* **914**, A5.

Deshpande, R., de Silva, C. M. & Marusic, I. 2023*a* Evidence that superstructures comprise self-similar coherent motions in high Reynolds number boundary layers. *J. Fluid Mech.* **969**, A10.

Deshpande, R., Van Den Bogaard, A., Vinuesa, R., Lindić, L. & Marusic, I. 2023*b* Reynolds-number effects on the outer region of adverse-pressure-gradient turbulent boundary layers. *Phys. Rev. Fluids* **8** (12), 124604.

Deshpande, R. & Vinuesa, R. 2024 Streamwise energy-transfer mechanisms in zero- and adverse-pressure-gradient turbulent boundary layers. *J. Fluid Mech.* **997**, A16.

Devenport, W. J. & Lowe, K. T. 2022 Equilibrium and non-equilibrium turbulent boundary layers. *Prog. Aerosp. Sci.* **131**, 100807.

Elsberry, K., Loeffler, J., Zhou, M. D. & Wygnanski, I. 2000 An experimental study of a boundary layer that is maintained on the verge of separation. *J. Fluid Mech.* **423**, 227–261.

Gul, M., Elsinga, G.E. & Westerweel, J. 2020 Internal shear layers and edges of uniform momentum zones in a turbulent pipe flow. *J. Fluid Mech.* **901**, A10.

Gungor, A. G., Maciel, Y., Simens, M. P. & Soria, J. 2016 Scaling and statistics of large-defect adverse-pressure-gradient turbulent boundary layers. *Int. J. Heat Fluid Flow* **59**, 109–124.

Gungor, T. R., Maciel, Y. & Gungor, A. G. 2022 Energy transfer mechanisms in adverse pressure gradient turbulent boundary layers: production and inter-component redistribution. *J. Fluid Mech.* **948**, A5.

Ho, C.-M. & Huerre, P. 1984 Perturbed free shear layers. *Annu. Rev. Fluid Mech.* **16**, 365–424.

Hutchins, N. & Marusic, I. 2007 Evidence of very long meandering features in the logarithmic region of turbulent boundary layers. *J. Fluid Mech.* **579**, 1–28.

Jiménez, J. 2018 Coherent structures in wall-bounded turbulence. *J. Fluid Mech.* **842**, P1.

Jones, M. B., Marusic, I. & Perry, A. E. 2001 Evolution and structure of sink-flow turbulent boundary layers. *J. Fluid Mech.* **428**, 1–27.

Kitsios, V., Sekimoto, A., Atkinson, C., Sillero, J. A., Borrell, G., Gungor, A. G., Jiménez, J. & Soria, J. 2017 Direct numerical simulation of a self-similar adverse pressure gradient turbulent boundary layer at the verge of separation. *J. Fluid Mech.* **829**, 392–419.

Kline, S. J., Reynolds, W. C., Schraub, F. A. & Runstadler, P. W. 1967 The structure of turbulent boundary layers. *J. Fluid Mech.* **30** (4), 741–773.

Knopp, T., Reuther, N., Novara, M., Schanz, D., Schülein, E., Schröder, A. & Kähler, C. J. 2021 Experimental analysis of the log law at adverse pressure gradient. *J. Fluid Mech.* **918**, A17.

Lee, J., Lee, J. H., Lee, J.-H. & Sung, H. J. 2010 Coherent structures in turbulent boundary layers with adverse pressure gradients. *J. Turbul.* **11**, N28.

Lee, J. H. 2017 Large-scale motions in turbulent boundary layers subjected to adverse pressure gradients. *J. Fluid Mech.* **810**, 323–361.

Lee, J.-H. & Sung, H. J. 2009 Structures in turbulent boundary layers subjected to adverse pressure gradients. *J. Fluid Mech.* **639**, 101–131.

Lindić, L., Abu Rowin, W., Deshpande, R. & Marusic, I. 2025 Investigation of turbulent/non-turbulent interfaces in high Reynolds number adverse pressure gradient boundary layers. *Int. J. Heat Fluid Flow* **114**, 109815.

Lozier, M., Deshpande, R., Zarei, A., Lindić, L., Rowin, W. A. & Marusic, I. 2025 Defining the mean turbulent boundary layer thickness based on streamwise velocity skewness. *J. Fluid Mech.* **1021**, A19.

Lozier, M., Thomas, F. O. & Gordeyev, S. 2026*a* Response of a turbulent boundary layer to a synthetic periodic large-scale structure. *arXiv preprint arXiv:2606.18602* .

LOZIER, M., ZAREI, A., MARUSIC, I. & DESHPANDE, R. 2026*b* On the wake region of high-Reynolds-number turbulent boundary layers subject to adverse pressure gradients. *Proceedings of the Fourteenth International Symposium on Turbulence and Shear Flow Phenomena, In Press* (179), arXiv preprint arXiv:2605.21872.

MARUSIC, I., CHAUHAN, K. A., KULANDAIVELU, V. & HUTCHINS, N. 2015 Evolution of zero-pressure-gradient boundary layers from different tripping conditions. *J. Fluid Mech.* **783**, 379–411.

MARUSIC, I. & HEUER, W. D. C. 2007 Reynolds number invariance of the structure inclination angle in wall turbulence. *Phys. Rev. Lett.* **99** (11), 114504.

MCKEON, B. J. 2017 The engine behind (wall) turbulence: perspectives on scale interactions. *J. Fluid Mech.* **817**, P1.

MONKEWITZ, P. A., CHAUHAN, K. A. & NAGIB, H. M. 2007 Self-consistent high-Reynolds-number asymptotics for zero-pressure-gradient turbulent boundary layers. *Phys. Fluids* **19** (11).

NAGANO, Y., TSUJI, T. & HOURA, T. 1998 Structure of turbulent boundary layer subjected to adverse pressure gradient. *Int. J. Heat Fluid Flow* **19** (5), 563–572.

NICKELS, T. B. 2004 Inner scaling for wall-bounded flows subject to large pressure gradients. *J. Fluid Mech.* **521**, 217–239.

PARTHASARATHY, A. & SAXTON-FOX, T. 2025 Turbulent boundary layers under spatially and temporally varying pressure gradients. *J. Fluid Mech.* **1010**, A61.

POZUELO, R., LI, Q., SCHLATTER, P. & VINUESA, R. 2022 An adverse-pressure-gradient turbulent boundary layer with nearly constant $\beta \simeq 1.4$ up to $Re_\theta \simeq 8700$. *J. Fluid Mech.* **939**, A34.

PRESKETT, T. & GANAPATHISUBRAMANI, B. 2025 The impact of pressure gradient history on flow structures in high Reynolds number rough-wall turbulence. *Int. J. Heat Fluid Flow* **116**, 109942.

ROMERO, S., ZIMMERMAN, S., PHILIP, J., WHITE, C. & KLEWICKI, J. 2022 Properties of the inertial sublayer in adverse pressure-gradient turbulent boundary layers. *J. Fluid Mech.* **937**, A30.

SCHATZMAN, D. M. & THOMAS, F. O. 2017 An experimental investigation of an unsteady adverse pressure gradient turbulent boundary layer: embedded shear layer scaling. *J. Fluid Mech.* **815**, 592–642.

SCHLATTER, P. & ÖRLÜ, R. 2012 Turbulent boundary layers at moderate Reynolds numbers: inflow length and tripping effects. *J. Fluid Mech.* **710**, 5–34.

SILVA, L. J. O. & WOLF, W. R. 2024 Embedded shear layers in turbulent boundary layers of a NACA0012 airfoil at high angles of attack. *Int. J. Heat Fluid Flow* **107**, 109353.

SMITS, A. J., MCKEON, B. J. & MARUSIC, I. 2011 High–Reynolds-number wall turbulence. *Annu. Rev. Fluid Mech.* **43** (1), 353–375.

TANARRO, Á., VINUESA, R. & SCHLATTER, P. 2020 Effect of adverse pressure gradients on turbulent wing boundary layers. *J. Fluid Mech.* **883**, A8.

TINNEY, C. E., COIFFET, F., DELVILLE, J., HALL, A. M., JORDAN, P. & GLAUSER, M. N. 2006 On spectral linear stochastic estimation. *Exp. Fluids* **41** (5), 763–775.

VINUESA, R., NEGI, P.S., ATZORI, M., HANIFI, A., HENNINGSON, D.S. & SCHLATTER, P. 2018 Turbulent boundary layers around wing sections up to $Re_c$ = 1,000,000. *Int. J. Heat Fluid Flow* **72**, 86–99.

VOLINO, R. J., SCHULTZ, M. P. & FLACK, K. A. 2007 Turbulence structure in rough- and smooth-wall boundary layers. *J. Fluid Mech.* **592**, 263–293.

WEI, T., LI, Z., KNOPP, T. & VINUESA, R. 2023 The mean wall-normal velocity in turbulent boundary layer flows under pressure gradient. *J. Fluid Mech.* **975**, A27.

WU, Y. & CHRISTENSEN, K. T. 2006 Population trends of spanwise vortices in wall turbulence. *J. Fluid Mech.* **568**, 55–76.

YANG, X.I.A., HU, R., DESHPANDE, R., KUNZ, R. & HUANG, G. 2026 Physical meaning of $k$ in the logarithmic layer for reynolds-averaged navier-stokes models. *Phys. Rev. Fluids* **11** (6), 064606.

YOON, M., HWANG, J., YANG, J. & SUNG, H.J. 2020 Wall-attached structures of streamwise velocity fluctuations in an adverse-pressure-gradient turbulent boundary layer. *J. Fluid Mech.* **885**, A12.

ZAREI, A., LOZIER, M., DESHPANDE, R. & MARUSIC, I. 2026*a* High-Reynolds-number turbulent boundary layers under adverse pressure gradients. Part 1. Decoupling local and upstream pressure gradient effects. *accepted in J. Fluid Mech. (in press)* **[arXiv:2509.07545v2]**.

ZAREI, A., LOZIER, M., DESHPANDE, R. & MARUSIC, I. 2026*b* High-Reynolds-number turbulent boundary layers under adverse pressure gradients. Part 2. A composite mean velocity profile. *accepted in J. Fluid Mech. (in press)* **[arXiv:2603.23912]**.

ZHOU, J., ADRIAN, R. J., BALACHANDAR, S. & KENDALL, T. M. 1999 Mechanisms for generating coherent packets of hairpin vortices in channel flow. *J. Fluid Mech.* **387**, 353–396.